\documentclass[prx,preprint,superscriptaddress]{revtex4}

\usepackage{amsmath}
\usepackage{amsfonts}
\usepackage{amssymb}
\usepackage{bm}
\usepackage{txfonts}
\usepackage[dvips]{graphicx}
\usepackage{multirow}
\usepackage{threeparttable}
\usepackage[usenames]{color}
\usepackage{ulem}
\usepackage{subfigure}
\def\bk{\bm k}
\def\br{\bm r}

\def\d{\partial }

\def\e{\varepsilon }

\def\g{\gamma }

\def\lm{\lambda }
\def\om{\omega }

\def\R{\mbox{\boldmath $R$} }

\def\G{\mbox{\boldmath $G$} }

\def\A{\mbox{\boldmath $A$} }

\def\j{\mbox{\boldmath $j$} }

\def\bmu{\mbox{\boldmath $\mu$} }
 
\def\AA{$\mathrm{\mathring{A}}$} 

\def\Tr{\mathop{\rm Tr}\nolimits}

\def\bra#1{\langle\,{#1}\,\vert\,}
\def\ket#1{\,\vert\,{#1}\,\rangle}
\def\braket#1#2{\langle\,{#1}\,\vert\,{#2}\,\rangle}

\def\tilde{\widetilde}
\def\lra{$\leftrightarrow$ }
\def\hs{\hskip3mm }

\begin{document}

\newcounter{ENVeqcntsave}
\newenvironment{ENV}

%\preprint{}

%\title{Exact extended quasiparticle theory of excited-state
% statics and dynamics of materials}
\title{
A simple derivation of the exact quasiparticle theory and
its extension to arbitrary initial excited eigenstates}


\author{Kaoru Ohno}
\email[]{ohno@ynu.ac.jp}
\affiliation{
Department of Physics, Yokohama National University,
79-5 Tokiwadai, Hodogaya-ku, Yokohama 240-8501, Japan
}
\author{Shota Ono}
\affiliation{
Department of Physics, Yokohama National University,
79-5 Tokiwadai, Hodogaya-ku, Yokohama 240-8501, Japan
}
\affiliation{
Department of Electrical, Electronic and Computer Engineering, Gifu University,
1-1 Yanagido, Gifu City 501-1193, Japan
}
%\author{Thi Nu Pham}
%\affiliation{
%Department of Physics, Yokohama National University,
%79-5 Tokiwadai, Hodogaya-ku, Yokohama 240-8501, Japan
%}
\author{Tomoharu Isobe}
\affiliation{
Department of Physics, Yokohama National University,
79-5 Tokiwadai, Hodogaya-ku, Yokohama 240-8501, Japan
}
%\thanks{}


\date{\today}

\begin{abstract}
The quasiparticle (QP) energies, which are minus of the energies
required by removing or produced by adding one electron from/to the system,
corresponding to the photoemission or inverse photoemission (PE/IPE) spectra,
are determined together with the QP wave functions, which are not orthonormal
and even not linearly independent but somewhat similar to the normal spin orbitals
in the theory of the configuration interaction,
by self-consistently solving the QP equation coupled with the equation for the self-energy.
The electron density, kinetic and all interaction energies can be calculated using the QP wave functions.
We prove in a simple way that the PE/IPE spectroscopy and therefore this QP theory can be applied
to an arbitrary initial excited eigenstate.
In this proof, we show that the energy-dependence of the self-energy is not an essential difficulty,
and the QP picture holds exactly if there is no relaxation mechanism in the system.
The validity of the present theory for some initial excited eigenstates is tested
using the one-shot GW approximation for several atoms and molecules.
\end{abstract}

%\pacs{78.47.da, 71.15.-m, 31.70.Hq, 31.15.ee}
%\keywords{}

\maketitle

%\subsection{}
%\subsubsection{}

\section{Introduction}

There are variety of excited states of atoms, molecules and any kind of materials, for example,
irradiated by laser-, electron-, and ion-beams, and they show variety of fascinating behaviors.
However, lack of a first-principles theory, which can treat arbitrary electronic excited states of materials,
has strongly prevented the progress of theoretical studies of such excited states of materials.
In 1964, Hohenberg and Kohn \cite{HohenbergKohn} invented density functional theory (DFT)
on the basis of the variational principle in quantum mechanics.
Next year, adopting DFT in the local density approximation (LDA),
Kohn and Sham \cite{KohnSham} proposed to use the so-called Kohn--Sham (KS) equation,
which has a simple form of a single particle Schr\"{o}dinger equation with the Hartree and exchange-correlation potentials.
If one could solve the KS equation for continuous fractional occupation numbers $f_{\mu}$ ($f_{\nu}$)
for certain occupied (empty) KS levels $\mu$ ($\nu$),
one might obtain the quasiparticle (QP) energies, $E^N_{f_{\mu}=1}-E^{N-1}_{f_{\mu}=0}=\int_0^1\e_{\mu}(f_{\mu})df_{\mu}$
and $E^{N+1}_{f_{\nu}=1}-E^N_{f_{\nu}=0}=\int_0^1\e_{\nu}(f_{\nu})df_{\nu}$,
which should rigorously correspond to the experimental photoemission and inverse photoemission (PE/IPE) spectra,
thanks to Janak's theorem, $\e_{\lm}(f_{\lm}) =\d E/\d f_{\lm}$ for $\lm = \mu$ or $\nu$, by integration \cite{Janak}.
Moreover, in 1985, Almbladh and von Barth \cite{Almbladh} demonstrated that the correct treatment
of the exchange-correlation energy produces the KS eigenvalue for the HOMO (highest-occupied molecular orbital) level
that is identical to the minus of the experimental, {\it i.e.}, exact, ionization potential (IP).
However, even if one could perform such calculations, the resulting information is just the usual PE/IPE energy
or spectra of the system at the ground state only.

In 1984, in order to treat the non-steady states and the electronic excited states of materials,
the time-dependent density functional theory (TDDFT) was invented by Runge and Gross \cite{RungeGross}.
Although this is a very important theory that may treat arbitrary excited states of materials,
there is a serious problem that no systematic way is given to determine the exchange-correlation kernel $\hat{\mu}_{xc}$,
which is a functional of a whole temporal history of the electron density $\{\rho(\br,t)\}$
and the initial wave function $\Psi(\{\br_i\},t_0)$ at $t=t_0$.
So, one usually adopts the adiabatic LDA (A-LDA),
in which the whole temporal history is completely ignored in the time domain and the simplest LDA is used
in the space domain in the functional.
The reliability of the A-LDA has been discussed by many authors \cite{Dobson,Vignale,Kohn,Hessler,D'Amico,LevyNagy,Ohno}.
Recently, it has been also emphasized to use accurate exchange-correlation functional
in the study of excited states \cite{CrawfordUrangaRubio}.

Of course some of the electronic excited states may be treated with highly accurate quantum chemistry (QC) methods \cite{Sabo,CC}
such as the configuration interaction (CI), complete active space (CAS), M{\o}ller--Plesset (MP), and coupled cluster (CC) methods.
However, these wave function-based methods beyond the Hartree--Fock approximation (HFA) are computationally very heavy
and again restricted to the singly, doubly, or triply (perhaps at most quadruply) excited neutral/charged states only.

On the other hand, the Green's function method on the basis of the many-body perturbation theory (MBPT)
can directly describe the PE/IPE energy spectra, which are usually called the quasiparticle (QP) energies.
The number of the papers using this method such as the $GW$ approximation has increased recent years.
One difficulty of using this method is that the MBPT and the Green's function are not easy to understand
for the researchers who are not familiar to the field theory.
Therefore, the Green's function is not at all popular in the fields of quantum chemistry and materials science.
Moreover, there is a serious problem related to the energy-dependence of the self-energy,
which has been a long standing issue and the QP picture has sometimes been considered as an approximation.
And again, the main issue here is that the Green's function has been always defined on the neutral ground state only,
while the initial electronic state in the PE/IPE spectroscopy can
be any of the neutral or charged excited states.

The purpose of this paper is to present a simple derivation of the QP theory,
which is applicable to any of the initial electronic excited eigenstates $\ket{\Psi_{\g}^M}$
of the $M$-electron system ($M$ can be different from the total number of protons $N$ in the system)
without invoking the complicated field theoretical MBPT.
In the course of this derivation, we will show that the energy-dependence of the self-energy
is not an essential difficulty, and the QP picture holds rigorously if there is no relaxation mechanism such as
electron-photon and electron-phonon interactions or cascade damping in the continuum energy bands.
(Even when there is such a relaxation mechanism in the system,
the QP picture should still hold within the lifetime \cite{Abrikosov,Lifshitz},
or equivalently within the mean free path.)

Moreover, it becomes clear that the QP wave functions $\phi_{\lm}(\br,s)$,
which are defined as the overlap between the initial state $\ket{\Psi_{\g}^M}$
operated by a creation or annihilation operator ($\psi^\dagger_s(\br)$ or $\psi_s(\br)$)
and the final state $\bra{\Psi_{\lm}^{M\pm 1}}$ in a PE/IPE process, 
satisfy the completeness condition but are generally non-orthonormal and linearly dependent.
(A relation to the normal spin orbitals in the CI theory \cite{Sabo,Lowdin1} will be mentioned in Section \ref{theory}.) 
Although the QP theory looks like a one-electron approximation, it is not an approximation at all.
It should be emphasized that the QP energies and the QP wave functions rigorously include
all the necessary many-body information and formally exact in a sense that these quantities yield
exact electron density as well as the exact kinetic, local potential, and electron-electron interaction energies.
The present theory is exact for any initial excited eigenstate.

The rest of this paper is organized as follows.
In Section \ref{theory},
the QP energies and the QP wave functions are defined in connection to the PE/IPE spectroscopy,
and the equations satisfied by them are derived.
The exact expressions for the electron density and the expectation values for the kinetic,
local potential, and interaction energies are derived.
In Section \ref{Green}, the same equations are rederived using the Green's functions.
In Section \ref{Perturbation},
a concrete method to calculate the self-energy is derived by invoking
Brillouin--Winger's perturbation theory.
In Section \ref{test1}, several test calculations of isolated atoms, ions and molecules using the one-shot $GW$ approximation
demonstrate the validity of the present theory.
Section \ref{remarks} presents several important remarks:
The virial theorem, the Ward--Takahashi identity, which is equivalent to the local charge conservation
(the continuity equation), and the macroscopic conservation laws hold,
and the Luttinger--Ward functional exists for the total energy, for the excited eigenstate.
Finally the DFT universal functional of the density is extended to the excited eigenstate,
for which the occupation numbers of all KS levels become its additional dependence,
and a variational principle still holds in a restricted Fock space that are orthogonal
to all lower lying excited eigenstates.
Finally, Section \ref{summary} is devoted to summarizing this paper.

\section{Basic theory}
\label{theory}

The present formulation can be simply extended to infinite crystal systems,
although the summation over $\bk$ is not written explicitly here.
The applicability to crystals is obvious in the case of the ground state
because there have been so many papers using GW approximation for crystals.
How about the case of the excited states?
If only some finite number of electrons are excited from the ground state,
nothing changes in the infinite system.
Therefore, the calculations are only meaningful when the finite density of electrons are excited from the ground state.
Such calculations would correspond to strong excitations induced by e.g. laser irradiations.

\begin{figure}[hbtp]
\begin{center}
\includegraphics[width=50mm]{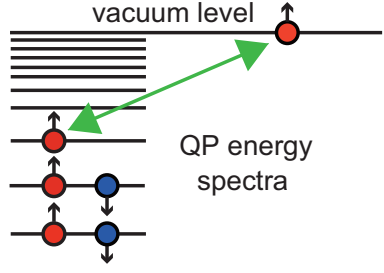}
\caption{Meaning of the QP energies}
\label{QPES}
\end{center}
\end{figure}

We start from the definition of the QPs. The initial state can be a ground
or any excited state of the $N$-electron neutral system or any excited state of
the $M(\neq N)$-electron charged system.
We write this state as $\ket{\Psi_{\g}^M}$ and its total energy as $E_{\g}^M$.
They satisfy the eigenvalue equation $H\ket{\Psi_{\g}^M} = E_{\g}^M\ket{\Psi_{\g}^M}$.
The electrons are interacting each other by the Coulomb interaction,
and therefore cannot be treated independently.
However, if we introduce the QP picture, we have a simple way to treat $\ket{\Psi_{\g}^M}$
in terms of the QPs.
The idea is based on Einstein's photoelectronic effect, which is called in modern terminology
the potoelectron or PE/IPE spectroscopy.
After the photoemission process, the system becomes a $(M-1)$-electron excited state $\ket{\Psi_{\mu}^{M-1}}$,
which satisfies the eigenvalue equation $H\ket{\Psi_{\mu}^{M-1}} = E_{\mu}^{M-1}\ket{\Psi_{\mu}^{M-1}}$.
The corresponding total energy $E_{\mu}^{M-1}$ satisfies
\begin{subequations}
\begin{align}
E_{\mu}^{M-1} - E_{\g}^M = h\nu - K
\end{align}
according to the energy conservation law, where $h\nu$ is the energy of the incident photon
and $K$ is the kinetic energy of the emitted electron.
Alternatively, after the inverse photoemission process, the system becomes a $(M+1)$-electron
excited state, which satisfy $H\ket{\Psi_{\nu}^{M+1}} = E_{\nu}^{M+1}\ket{\Psi_{\nu}^{M+1}}$.
In this case the total energies satisfy
\begin{align}
E_{\g}^M - E_{\nu}^{M+1}  = h\nu - K.
\end{align}
\end{subequations}
Therefore, the total energy differences are the measurable quantities.
We call these energies the QP energies and define them as
\begin{subequations}
\begin{align}
\e_{\mu} = E_{\g}^M - E_{\mu}^{M-1},
\label{qpe1}
\\
\e_{\nu} = E_{\nu}^{M+1} - E_{\g}^M.
\label{qpe2}
\end{align}
\label{qpe}
\end{subequations}
The QP energy $\e_{\mu}$ corresponds to the minus of the energy required to remove
one electron from the $\ket{\Psi_{\g}^M}$ state and add it at at the vacuum level; see Fig. \ref{QPES}.
Similarly, the QP energy $\e_{\mu}$ corresponds to the minus of the energy gain when
one electron is added to the $\ket{\Psi_{\g}^M}$ state from the vacuum level.
Note that the final states includes many ($M\pm 1$)-electron states,
which can be created by adding or removing one electron from the initial state.
This does not necessarily mean that the final states include all possible ($M\pm 1$)-electron states.
For example, multiple scattered eigenstates where many electron configurations are different from
the initial eigenstate are excluded or their amplitudes are quite small even if they are included.

\begin{figure}[hbtp]
\begin{center}
\subfigure[]{
    \includegraphics[width=0.3\textwidth]{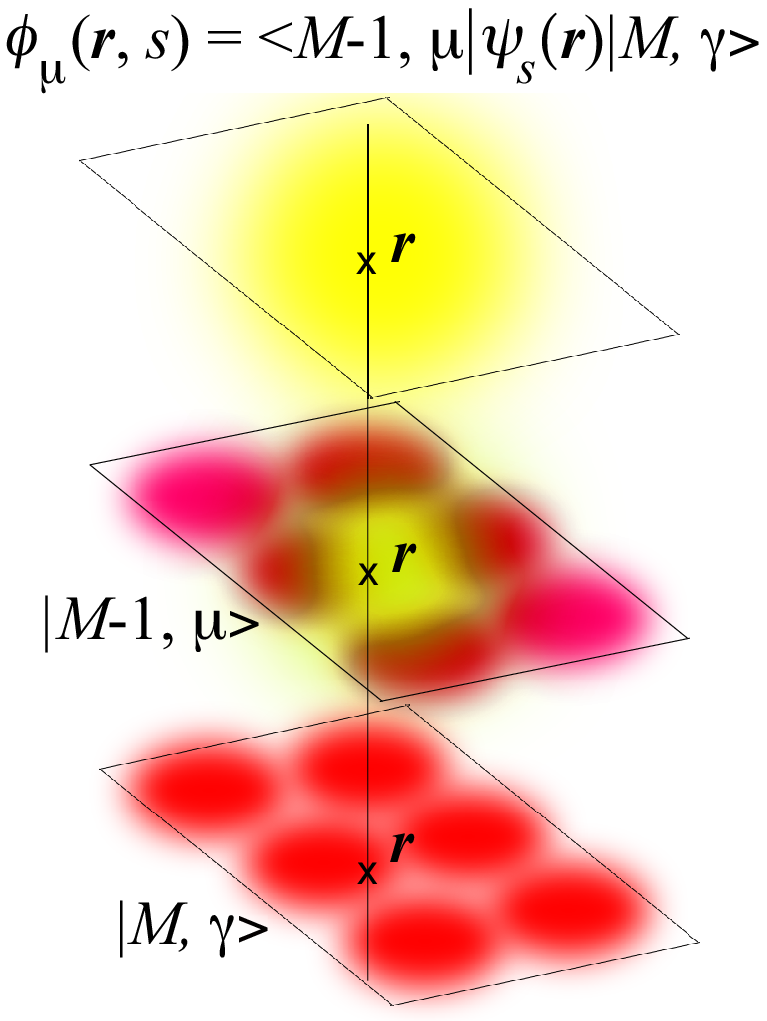}}
    \hskip10mm
\subfigure[]{
    \includegraphics[width=0.3\textwidth]{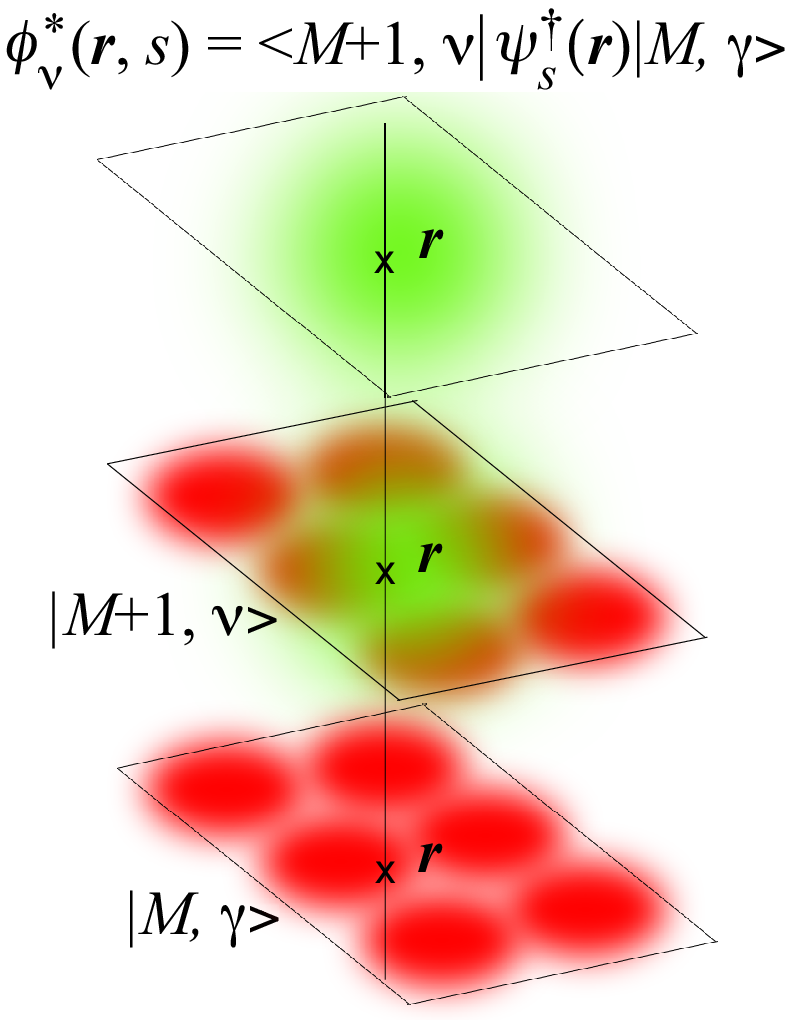}}
\caption{Meaning of the QP wave functions (overlap amplitudes).
Yellow and green parts represent the hole and electron distributions, respectively.}
\label{fig:QPWF}
\end{center}
\end{figure}

The QP wave functions can be defined as an overlap between the initial state 
$\ket{\Psi_{\g}^M}$ with one-electron removed or added at a point $(\br,s)$ 
( $\br$ and {\it s} are spacial and spin coordinates of the electron) 
and the final state $\ket{\Psi^{M-1}_\mu}$ or $\ket{\Psi^{M+1}_\nu}$
(therefore, they are sometimes called the overlap amplitudes):
\begin{subequations}
\begin{align}
\phi_\mu (\br,s) &= \bra{\Psi^{M-1}_\mu}  \psi_s (\br) \ket{ \Psi_{\g}^M},
\label{qpwf1}
\\
\phi^\ast_\nu (\br,s) &= \bra{\Psi^{M+1}_\nu } \psi^\dagger_s(\br) \ket{ \Psi_{\g}^M}.
\label{qpwf2}
\end{align}
\label{qpwf}
\end{subequations}
Here $\psi^\dagger_s(\br)$ and $\psi_s(\br)$ are the (creation and annihilation) field operators.
These QP wave functions represent the amplitude of the one-electron difference 
between the initial and the final states (see Fig. \ref{fig:QPWF}). 
The total electron spin density of the $\ket{\Psi_{\g}^M}$ state is given by
\begin{align}
\rho_s(\br) % & = \bra{\Psi_{\g}^M} \hat{n}_s(\br) \ket{ \Psi_{\g}^M}
%\nonumber\\
& = \bra{\Psi_{\g}^M} \psi^\dagger_s(\br) \psi_s(\br) \ket{\Psi_{\g}^M}
\nonumber \\ 
& =  \sum_\mu^{\rm occ} \bra{\Psi_{\g}^M} \psi^\dagger_s(\br) \ket{\Psi^{M-1}_\mu} 
\bra{\Psi^{M-1}_\mu} \psi_s(\br)  \ket{\Psi_{\g}^M} 
\nonumber \\ 
& = \sum_\mu^{\rm occ} \left| \phi_\mu(\br,s) \right|^2,
\label{SpinDensity}
\end{align}
where the completeness condition for the eigenstates $\ket{\Psi_\mu^{M-1}}$
of the ($M-1$)-electron Hamiltonian
\begin{align}
\sum_\mu^{\rm occ} \ket{\Psi^{M-1}_\mu} \bra{\Psi^{M-1}_\mu} = 1
\label{completeness1}
\end{align}
was used.
Similarly, the expectation value of the one-body part of the Hamiltonian
\begin{align}
H^{(1)} = \hat{T} + \hat{v} = \sum_s \int \psi_s^\dagger(\br) h_s^{(1)}(\br) 
\psi_s(\br) d\br, \ \ h_s^{(1)}(\br) = -\frac{\hbar^2}{2m} \nabla^2 + v(\br)
\label{H_1body}
\end{align}
($v(\br)$ is the local potential such as the Coulomb potential caused by nuclei)
are obtained by the QP wave functions as follows:
\begin{align}
\langle\hat{T}\rangle & = -\frac{\hbar^2}{2m} \bra{\Psi_{\g}^M} \sum_s \int
 \psi_{s}^\dagger (\br)  \nabla^2 \psi_{s} (\br) d\br \ket{\Psi_{\g}^M}
\nonumber \\
& = -\frac{\hbar^2}{2m} \sum_{\mu}^{\rm occ}\sum_s \int
\phi^\ast_\mu (\br,s) \nabla^2 \phi_\mu (\br,s) d\br,
\label{kineticE} \\
\langle\hat{v}\rangle & = \bra{\Psi_{\g}^M} \sum_s \int
 \psi_{s}^\dagger (\br)  v(\br) \psi_{s} (\br) d\br \ket{\Psi_{\g}^M}
\nonumber \\
& = \sum_{\mu}^{\rm occ}\sum_s \int v(\br) \left|\phi_\mu (\br,s)\right|^2 d\br.
\label{potentialE}
\end{align}
In fact, the QP wave functions diagonalize the density matrix
\begin{align}
\rho_s(\br',\br) = \bra{\Psi_{\g}^M}\psi^\dagger_s(\br')\psi_s(\br)\ket{\Psi_{\g}^M}
= \sum_{\mu}^{\rm occ} \,\phi^\ast_{\mu}(\br',s)\phi_{\mu}(\br,s).
\label{DensityMatrix}
\end{align}
The spin density $\rho_s(\br)$ is given by $\rho_s(\br,\br)$,
and the expectation values of the kinetic and local potential energies in the $\ket{\Psi_{\g}^M}$ state are given by
\begin{align}
\langle\hat{T}\rangle & = -\frac{\hbar^2}{2m} \sum_s \int \lim_{\br'\rightarrow\br} \nabla^2 \rho_s(\br',\br) d\br,
\label{kineticE2} \\
\langle\hat{v}\rangle & = \sum_s \int v(\br) \rho_s(\br,\br) d\br.
\label{potentialE2}
\end{align}
In this sense, the QP wave functions $\phi_{\mu}(\br,s)$ have a property similar to
the natural spin orbitals introduced by L\"{o}wdin more than 60 years ago \cite{Lowdin1,Lowdin2}
in connection with the CI theory \cite{Sabo}.

Here, it is beneficial to note that the completeness condition
\begin{align}
&\sum_\mu^{\rm occ} \phi_\mu(\br,s) \phi_\mu^\ast (\br',s') +
\sum_\nu^{\rm emp} \phi_\nu(\br,s) \phi_\nu^\ast (\br',s')
\nonumber\\
& = \sum_{\lm}^{\rm all} \phi_{\lm}(\br,s) \phi_{\lm}^\ast (\br',s') = \delta(\br - \br') \delta_{ss'}.
\label{complete}
\end{align}
holds for the QP states. This can be easily seen by sandwiching the anticommutation relation of the field operators
\begin{align}
\left\{\psi_s(\br), \psi^\dagger_{s'}(\br')\right\} = \delta(\br-\br')\delta_{ss'},
\label{anticomm}
\end{align}
with $\bra{\Psi_{\g}^M}$ and $\ket{\Psi_{\g}^M}$ and using
the other completeness condition for the eigenstates $\ket{\Psi_{\nu}^{M+1}}$ of the ($M+1$)-electron Hamiltonian
\begin{align}
\sum_{\nu}^{\rm emp}\ket{\Psi_\nu^{M+1}}\bra{\Psi_\nu^{M+1}}=1
\label{completeness2}
\end{align}
together with (\ref{completeness1}).
On the other hand, the QP wave functions are not orthonormal, {\it i.e.}, not normalized to unity,
and even not linearly independent \cite{HedinLundqvist,Goscinski},
and therefore they are different from the normal spin orbitals mentioned above in an exact sense.
However, one should notice that
the overlap matrix $S$ between the QP wave functions satisfies the idempotency relation $S = S^2$, {\it i.e.},
\begin{align}
S_{\lm\lm'} & = \sum_s\int\phi^\ast_{\lm}(\br,s)\phi_{\lm'}(\br,s)d\br
\nonumber\\
& = \sum_{ss'}\int\phi^\ast_{\lm}(\br,s)\delta(\br - \br') \delta_{ss'}\phi_{\lm'}(\br',s')d\br d\br'
\nonumber\\
& = \sum_{ss'}\int\phi^\ast_{\lm}(\br,s)
\sum_{\lm''}^{\rm all}\phi_{\lm''}(\br,s)\phi^\ast_{\lm''}(\br',s') \;
\phi_{\lm'}(\br',s')d\br d\br'
\nonumber\\
& = \sum_{\lm''}^{\rm all}
\sum_{s}\int\phi^\ast_{\lm}(\br,s)\phi_{\lm''}(\br,s)d\br
\sum_{s'}\int\phi^\ast_{\lm''}(\br',s')\phi_{\lm'}(\br',s')d\br'
\nonumber\\
& = \sum_{\lm''}^{\rm all}S_{\lm\lm''}S_{\lm''\lm},
\label{qpe_S}
\end{align}
where the completeness condition (\ref{complete}) was used in the third equality.
So, if one diagonalizes the overlap matrix $S$ by a unitary transformation as $\tilde{S}=U^\dagger SU$
according to the canonical orthonormalization procedure introduced by L\"{o}wdin \cite{Lowdin1,Lowdin2,Goscinski},
it is realized that the resulting diagonal matrix $\tilde{S}$ satisfies $\tilde{S}(\tilde{S}-1)=0$
because the original matrix $S$ satisfies $S(S-1)=0$.
Therefore the diagonal element of $\tilde{S}$, {\it i.e.}, the eigenvalues of $S$ must be either zero or unity.
Thus the restricted set of the transformed functions
\begin{align}
\tilde{\phi}_{\alpha}(\br,s) = \sum_{\lm}^{\rm all}\phi_{\lm}(\br,s)U_{\lm\alpha},
\label{unitary}
\end{align}
which correspond to the eigenvalue unity only,
becomes linearly independent and satisfies the orthonormality condition
\begin{align}
\tilde{S}_{\alpha\alpha'} = \int\tilde{\phi}^\ast_{\alpha}(\br,s)\tilde{\phi}_{\alpha'}(\br,s)d\br = \delta_{\alpha\alpha'}.
\label{orthonormal}
\end{align}
This restricted set of the transformed functions satisfies the idempotency relation $\tilde{S}=\tilde{S}^2$.
Since $\tilde{S}^2$ is given by
\begin{align}
\sum_{\alpha''}^{\rm restrict}\tilde{S}_{\alpha\alpha''}\tilde{S}_{\alpha''\alpha}
&  = \sum_{\alpha''}^{\rm restrict}
\sum_{s}\int\tilde{\phi}^\ast_{\alpha}(\br,s)\tilde{\phi}_{\alpha''}(\br,s)d\br
\sum_{s'}\int\tilde{\phi}^\ast_{\alpha''}(\br',s')\tilde{\phi}_{\alpha'}(\br',s')d\br'
\nonumber\\
& = \sum_{ss'}\int\tilde{\phi}^\ast_{\alpha}(\br,s)
\sum_{\alpha''}^{\rm restrict}\tilde{\phi}_{\alpha''}(\br,s)\tilde{\phi}^\ast_{\alpha''}(\br',s') \;
\tilde{\phi}_{\alpha'}(\br',s')d\br d\br',
\label{qpe_S'}
\end{align}
which must be equal to (\ref{orthonormal}), the completeness condition
\begin{align}
\sum_{\lm} \tilde{\phi}_{\lm}(\br,s) \tilde{\phi}^\ast_{\lm}(\br',s') = \delta(\br - \br') \delta_{ss'}
\label{complete2}
\end{align}
is fulfilled also for the transformed functions \cite{Goscinski}.
In practical calculations, however, one usually assumes that the diagonal elements $S_{\lm\lm}$ are unity.
In such cases, one has to redetermine these norms afterwards, because they are generally not unity at all.
In this procedure, the unitary transformed $S'$ is diagonal but its diagonal elements are no more zero or unity.
It has a form \cite{Goscinski}
\begin{align}
%\tilde{S} = U'^\dagger SU' = \left(
S' = U'^\dagger SU' = \left(
\begin{array}{cc}
\bmu \; & \; 0 \\
0 \; & \; 0
\end{array}
\right),
\end{align}
where the submatrix $\bmu$ is a diagonal matrix, whose elements are not necessarily unity.
Then, the orthonormal set must be chosen as
\begin{align}
\tilde{\phi}_{\alpha}(\br,s) = \frac{1}{\sqrt{\bmu_{\alpha\alpha}}}\sum_{\lm}^{\rm all}\phi_{\lm}(\br,s)U'_{\lm\alpha}
\label{unitary2}
\end{align}
instead of (\ref{unitary}). From this equation,
the QP wave functions with the correct norm $S_{\lm\lm}$ are found to be
\begin{align}
\phi_{\lm}(\br,s) = \sqrt{S_{\lm\lm}}
\sum_{\alpha}^{\rm restrict}\sqrt{\bmu_{\alpha\alpha}}\tilde{\phi}_{\alpha}(\br,s)U'^\dagger_{\alpha\lm}.
\label{unitary3}
\end{align}
From Eq. (\ref{unitary}), they should be equal to
\begin{align}
\phi_{\lm}(\br,s) = \sum_{\alpha}^{\rm restrict}\tilde{\phi}_{\alpha}(\br,s)U^\dagger_{\alpha\lm}.
\label{unitary3}
\end{align}
Comparing these two equations, we have
\begin{align}
\sqrt{\bmu_{\alpha\alpha}}U'^\dagger_{\alpha\lm}\sqrt{S_{\lm\lm}}
 = U^\dagger_{\alpha\lm}.
\label{unitary3}
\end{align}
Because $U_{\alpha\lm}$ is unitary,
\begin{align}
\sum_{\alpha}^{\rm restrict}
U'_{\lm\alpha}\,\bmu_{\alpha\alpha}
U'^\dagger_{\alpha\lm'}
 = \left(1/S_{\lm\lm}\,\right)\delta_{\lm\lm'}
\label{unitary3}
\end{align}
is obtained.
Therefore, one can determine the norm of each QP wave function $S_{\lm\lm}$ using this procedure.
Apart from this complexity, the non-orthogonality and linear dependence of the QP wave functions
do not bring any essential difficulty in the present formulation.

What we want to emphasize here is that there are vast amount of irrelevant ($M\pm 1$)-electron states,
which can be ignored from the summation over all empty states or all occupied states
in Eqs. (\ref{SpinDensity})-(\ref{DensityMatrix}) and (\ref{complete}).
For example, one may consider a lot of independent excitations that are not directly associated with
the photoemission process under consideration.
%Even such excited states, they are the ($M-1$)-electron states,
Even such excited states are the ($M-1$)-electron states,
which can be included in the summation over all occupied states.
Of course, such states do not contribute to the intermediate states
in Eqs. (\ref{SpinDensity})-(\ref{DensityMatrix}) and (\ref{complete}),
because such multiple excitations cannot occur with the single creation or annihilation operator.
This is a clear example of the non-orthonormality: The amplitude for such QP states is zero or almost zero.
The simplest way is to simply remove such ($M\pm 1$)-electron states from the QP states.
(We will use this approximation in our test calculations given in Section \ref{test1}.)

Now, let us derive the equation satisfied by the QP wave functions and the QP energies.
Since the Hamiltonian $H$ is given by $H = H^{(1)} + H^{(2)}$ with the one-body part (\ref{H_1body})
and the two-body part, {\it i.e.}, the electron-electron Coulomb interaction,
\begin{align}
H^{(2)} = \frac{1}{2}\sum_{ss'} \int \psi_s^\dagger(\br)
\psi_{s'}^\dagger(\br') h_{ss'}^{(2)}(\br,\br') \psi_{s'}(\br') \psi_s(\br) d\br d\br',
\ \ h_{ss'}^{(2)}(\br,\br') = \frac{e^2}{4\pi\e_0} \frac{1}{ | \br - \br'|},
\label{H_2body}
\end{align}
the commutation relation between $\psi_s (\br)$ and $H$ is given by
\begin{align}
[\psi_s(\br), H] = h_s^{(1)}\psi_s(\br) + \sum_{s'} \int 
\psi_s^\dagger (\br') h_{s's}^{(2)} (\br',\br) \psi_{s'}(\br')  \psi_s(\br) d\br'.
\end{align}
Then, by sandwiching this with $\bra{\Psi^{M-1}_\mu}$ and $\ket{\Psi_{\g}^M}$
or with $\bra{\Psi_{\g}^M}$ and $\ket{\Psi^{M+1}_\nu}$,
and using Eqs. (\ref{qpe}) and (\ref{qpwf}), we derive
\begin{subequations}
\begin{align}
%\left( E_{\g}^M - E^{M-1}_\mu \right)
\e_{\mu} \phi_\mu (\br,s) = h_s^{(1)} (\br) 
\phi_\mu (\br,s) + \sum_{s'}\int h^{(2)}_{s's} (\br', \br ) 
\bra {\Psi^{M-1}_\mu}\psi_{s'}^\dagger (\br')\psi_{s'}(\br') \psi_s(\br)
\ket{\Psi_{\g}^M} d\br',
\label{way_of_QP1}
\\
\e_{\nu} \phi_\nu (\br,s) = h_s^{(1)} (\br) 
\phi_\nu (\br,s) + \sum_{s'}\int h^{(2)}_{s's} (\br', \br ) 
\bra {\Psi_{\g}^M}\psi_{s'}^\dagger (\br')\psi_{s'}(\br') \psi_s(\br)
\ket{\Psi_{\nu}^{M+1}} d\br',
\label{way_of_QP2}
\end{align}
\end{subequations}
Here, the second term in the right-hand side (r.h.s.) of these equations are rewritten as
\begin{subequations}
\begin{align}
&\sum_{s'} \int h^{(2)}_{s's} (\br', \br ) 
\bra {\Psi^{M-1}_\mu}\psi_{s'}^\dagger (\br')\psi_{s'}(\br') \psi_s(\br)\ket{\Psi_{\g}^M} d\br'
\nonumber\\
& = \int \Sigma_s (\br, \br', E_{\g}^M - E_{\mu}^{M-1})\bra {\Psi^{M-1}_\mu}\psi_{s}(\br')\ket{\Psi_{\g}^M}
 d\br',
\label{Sigma1a}
\end{align}
\begin{align}
& \sum_{s'} \int h^{(2)}_{s's} (\br', \br ) 
\bra {\Psi_{\g}^M}\psi_{s'}^\dagger (\br')\psi_{s'}(\br') \psi_s(\br)\ket{\Psi_{\nu}^{M+1}} d\br'
\nonumber\\
& = \int \Sigma_s (\br, \br', E_{\nu}^{M+1} - E_{\g}^M)\bra {\Psi_{\g}^M}\psi_{s}(\br')\ket{\Psi_{\nu}^{M+1}}
d\br',
\label{Sigma1b}
\end{align}
\label{Sigma1}
\end{subequations}
by introducing the energy-dependent self-energy $\Sigma_s(\br,\br'; \e_{\lm})$
with $\e_{\lm}=\e_{\mu}$ or $\e_{\nu}$ given by Eq. (\ref{qpe}).
(Here and hereafter, the self-energy includes the Hartree term, which describes the classical Coulomb interaction.)
It is clear that both identities (\ref{Sigma1a}) and (\ref{Sigma1b}), which define the self-energy,
are consistent with each other, in particular, in the excited-state formulation.
In fact, since $\bra{\Psi_{\mu}^{M-1}}$ and $\ket{\Psi_{\g}^M}$ are arbitrary excited eigenstates in Eq. (\ref{Sigma1a}),
they can be changed to $\bra{\Psi_{\g}^M}$ and $\ket{\Psi_{\nu}^{M+1}}$, and then Eq. (\ref{Sigma1b}) is readily obtained.
In other words, in order to have the two identities (\ref{Sigma1a}) and (\ref{Sigma1b}) simultaneously,
one must recognize that the ordinary formulation for the self-energy for the ground state should hold
for arbitrary excited eigenstates also.
%Inserting (\ref{qpe}) and (\ref{qpwf}) into (\ref{Sigma1}), and using the orthonormality of the QP wave functions, 
%we anticipate that the self-energy may be written as
%\begin{subequations}
%\begin{align}
%\Sigma_s (\br, \br', H^{\rm QP})
%& = \sum_{\mu}^{\rm occ} \phi^\ast_\mu(\br',s) \sum_{s''} \int h^{(2)}_{s''s} (\br'', \br ) 
%\bra {\Psi^{M-1}_\mu}\psi_{s''}^\dagger (\br'')\psi_{s''}(\br'') \psi_s(\br)\ket{\Psi_{\g}^M} d\br''
%\nonumber\\
%& + \sum_{\nu}^{\rm occ} \sum_{s''} \int h^{(2)}_{s''s} (\br'', \br ) 
%\bra {\Psi_{\g}^M}\psi_{s''}^\dagger (\br'')\psi_{s''}(\br'') \psi_s(\br)\ket{\Psi_{\nu}^{M+1}} d\br'' \phi^\ast_\nu(\br',s)
%\nonumber\\
%& = \sum_{s''} \int h^{(2)}_{s''s} (\br'', \br ) 
%\bra {\Psi^{M-1}_\mu}\psi_s^\dagger (\br')\psi_{s''}^\dagger (\br'')\psi_{s''}(\br'') \psi_s(\br)\ket{\Psi_{\g}^M} d\br''
%\nonumber\\
%& + \sum_{s''} \int h^{(2)}_{s''s} (\br'', \br ) 
%\bra {\Psi_{\g}^M}\psi_{s''}^\dagger (\br'')\psi_{s''}(\br'') \psi_s(\br)\psi_s^\dagger (\br')\ket{\Psi_{\nu}^{M+1}} d\br'',
%\label{Sigma2}
%\end{align}
%\end{subequations}
%where $H^{\rm QP}$ in $\Sigma_s$ is replaced by the QP energy $\e_{\lm}$ when it operates on the QP wave function
%$\phi_{\lm}(\br',s)$.

From (\ref{Sigma1a}) and (\ref{Sigma1b}), the QP equations
\begin{subequations}
\begin{align}
h_s^{(1)}(\br) \phi_\mu(\br,s) + \int \Sigma_s (\br, \br', \e_\mu) 
\phi_\mu (\br' , s ) d\br' = \e_\mu \phi_\mu (\br,s),
\label{QPeq1}
\\
h_s^{(1)}(\br) \phi_\nu(\br,s) + \int \Sigma_s (\br, \br', \e_\nu) 
\phi_\nu (\br' , s ) d\br' = \e_\nu \phi_\nu (\br,s)
\label{QPeq2}
\end{align}
\label{QPeq}
\end{subequations}
are derived.
Since there is the energy-dependence in the self-energy, we have to solve these equations level to level.
Note that the self-energy has a state dependence only through the dependence on a particular QP energy.
In the case of the ground state, it has been demonstrated that the energy dependence can be approximately treated
by linearization \cite{Kresse,Kuwahara,Kuwahara2}. However, the energy-dependence is very important in the QP theory.
Indeed, multiplying both sides of Eq. (\ref{Sigma1a}) by 
$\frac{1}{2}\bra{\Psi_{\g}^M}\psi^\dagger_s (\br)\ket{\Psi^{M-1}_\mu}=\frac{1}{2}\phi^\ast_\mu (\br,s)$,
summing up with respect to $\mu$ and $s$, integrating with respect to $\br$,
and using the completeness condition (\ref{completeness1}) again, we obtain
\begin{align}
E_{\rm int} & = \frac{1}{2}\sum_{ss'} \int h_{ss'}^{(2)}(\br,\br') 
\bra{\Psi_{\g}^M}
 \psi_s^\dagger(\br) \psi_{s'}^\dagger(\br') \psi_{s'}(\br') \psi_s(\br) 
\ket{\Psi_{\g}^M} d\br d\br'
\nonumber\\
& = \frac{1}{2}\sum_{\mu}^{\rm occ}\sum_s \int \phi^\ast_\mu (\br,s) \Sigma_s(\br,\br';\e_\mu) \phi_\mu (\br',s) d\br d\br'
\label{Eint}
\end{align}
for the expectation value of the electron-electron Coulomb interaction for the $\ket{\Psi_{\g}^M}$ state,
which is known as the Galitskii and Migdal's formula \cite{Galitskii}
\begin{align}
E_{\rm int} = - \frac{1}{2}\Tr\int\Sigma_s(\br,\br';\om)G_s(\br',\br;\om)d\br'
\label{Galitskii}
\end{align}
(Notations appearing here will be explained in Section \ref{remarks}.)
in the QP representation \cite{Sanchez-Friera} for the case of the ground state.
The crucial points here are that the energy dependence of the self-energy appears in the same way
in both Eqs. (\ref{QPeq1}) and (\ref{Eint}) in a general scheme of the non-orthonormal QP wave functions
and that Eq. (\ref{Eint}) holds not only for the ground state but also for arbitrary excited eigenstates.
Equation (\ref{Eint}) implies that the information of the self-energy is enough to determine
the total energy $E_{\g}^M$ of the $\ket{\Psi_{\g}^M}$ state of the system, because the QP wave functions
are uniquely obtained by solving the QP equation (\ref{QPeq}).
In other words, if the self-energy is known, the QP energies and QP wave functions can be calculated
by solving the QP equation (\ref{QPeq}) one by one, and then the electron spin density $\rho_s(\br)$,
the kinetic energy $\langle\hat{T}\rangle$, the local potential energy $\langle\hat{v}\rangle$,
and the interaction energy $E_{\rm int}$ are obtained via Eqs. (\ref{SpinDensity}),
(\ref{kineticE}), (\ref{potentialE}), and (\ref{Eint}), respectively.
Consequently the total energy $E_{\g}^M$ is obtained from the knowledge of the self-energy only.
How to determine the self-energy will be described in Section \ref{Perturbation}.
%In the next Section, we will show that the equations derived in Section \ref{theory}
In next section (Sec. \ref{Green}), we will show that the equations derived in Section \ref{theory}
can be rederived by using the Green's function.

\section{Relation to the Green's Function Formalism} 
\label{Green}

The equations (\ref{Sigma1}) and (\ref{QPeq}) can be also derived from
the Dyson equation for the Green's function as follows.
The Green's function $G$ on the eigenstate $\ket{\Psi_{\g}^M}$ is defined as
\begin{align}
G_s(x,x')
&=-i\,\bra{\Psi_{\g}^M}T[\psi_s(x)\psi^{\dagger}_s(x')]\ket{\Psi_{\g}^M},
\label{G1a}
\end{align}
where the convention to write $x=(\br,t)$ is used;
$T$ denotes the time-ordered product, and
$\psi^{\dagger}_s(x)=e^{iHt}\psi^{\dagger}_s(\br)e^{-iHt}$ and $\psi_s(x)=e^{iHt}\psi_s(\br)e^{-iHt}$
are, respectively, the creation and annihilation operators in the Heisenberg representation.
By means of the completeness conditions (\ref{completeness1}) and (\ref{completeness2}),
\begin{align}
G_s(x,x')
= & i\sum_{\mu}^{\rm occ}\phi_{\mu}(x,s)\,\phi^\ast_{\mu}(x',s)\,
\theta(t'-t)
\nonumber\\
- & i\sum_{\nu}^{\rm emp}
\phi_{\nu}(x,s)\,\phi^\ast_{\nu}(x',s)\,
\theta(t-t')
\label{G1b}
\end{align}
is readily derived. Here we put
\begin{subequations}
\begin{align}
\phi_\mu(x,s)=\bra{\Psi_{\mu}^{M-1}}\psi_s(x)\ket{\Psi_{\g}^M}=\phi_\mu(\br,s)e^{-i\e_\mu t},
\label{phiTD1}
\\
\phi_\nu(x,s)=\bra{\Psi_{\g}^M}\psi_s(x)\ket{\Psi_{\nu}^{M+1}}=\phi_\nu(\br,s)e^{-i\e_\nu t},
\label{phiTD2}
\end{align}
\label{phiTD}
\end{subequations}
and $\phi_{\lm}(\br,s)$ and $\e_{\lm}$ ($\lm=\mu$ or $\nu$) are the QP wave functions and the QP energies
defined in Eqs. (\ref{qpe}) and (\ref{qpwf}).

The Heisenberg equation of motion for the annihilation operator is given by
\begin{align}
&i\frac{\d}{\d t}\psi_s(x)=[\psi_s(x),H]=h^{(1)}(\br)\,\psi_s(x)
\nonumber\\
+&\sum_{s''}\int h^{(2)}_{s''s}(\br'',\br)
\left[\psi_{s''}^{\dagger}(x'')\psi_{s'}(x'')\right]_{t''=t}d\br''\,\psi_s(x).
\label{Heisenberg1}
\end{align}
Operating both sides $\psi_s^{\dagger}(x')$ from the right and the time-ordered operator $T$ from the left, 
and sandwiching them with $\bra{\Psi_{\g}^M}$ and $\ket{\Psi_{\g}^M}$, we have
\begin{align}
&\left[i\frac{\d}{\d t}-h_s^{(1)}(\br)\right] G_s(x,x')
= \delta(x-x')
\nonumber\\
& +i\sum_{s''} \int h^{(2)}_{s''s}(\br'',\br)\bra{\Psi_{\g}^M}T\bigl[
\psi_s(x)\psi_{s''}(x'')\psi_{s''}^{\dagger}(x''^+)\psi_s^{\dagger}(x')
\bigr]\ket{\Psi_{\g}^M}\bigr|_{t''=t}\, d\br''
\end{align}
where $x''^+$ with $t''=t$ means $(\br'',t^+)$ with $t^+=t+0^+$.
The self-energy is introduced as
\begin{align}
& \int\Sigma_s(x,x'')G_s(x'',x')dx''
\nonumber\\
& = i\sum_{s''} \int h^{(2)}_{s''s}(\br'',\br)\bra{\Psi_{\g}^M}T\bigl[
\psi_s(x)\psi_{s''}(x'')\psi_{s''}^{\dagger}(x''^+)\psi_s^{\dagger}(x')
\bigr]\ket{\Psi_{\g}^M}\bigr|_{t''=t}\, d\br'',
\label{S1}
\end{align}
and then the Dyson equation
\begin{align}
\left[i\frac{\d}{\d t} - h_s^{(1)}(\br)\right] G_s(x,x')
 = \delta(x-x')
 + \int\Sigma_s(x,x'')G_s(x'',x')dx''
 \label{Dyson}
\end{align}
holds for any eigenstate. Equation (\ref{Dyson}) is rewritten in the Fourier space as
\begin{align}
\left[\om - h_s^{(1)}(\br)\right] G_s(\br,\br';\om)
 = \delta(\br-\br'')
 + \int\Sigma_s(\br,\br'';\om)G_s(\br'',\br';\om)d\br''.
\label{Dyson_w}
\end{align}
The Fourier transform of the Green's function (\ref{G1b}) with (\ref{phiTD}) is given by
\begin{align}
G_s(\br,\br';\om)
= \sum_{\lm}^{\rm all}
\frac{\phi_{\lm}(\br,s)\,\phi^*_{\lm}(\br',s)}
{\om-\e_{\lm}-i\delta_{\lm}},
\label{Lehman}
\end{align}
where $\delta_{\lm}=0^+$ for occupied states and $\delta_{\lm}=-0^+$ for empty states,
{\it i.e.}, $\delta_{\mu}=0^+$ and $\delta_{\nu}=-0^+$.
Inserting this into Eq. (\ref{Dyson_w})
and approaching $\om$ to one of the $\e_{\lm}$'s
($\lm$ can be either $\nu$ (emp) or $\mu$ (occ)), 
we obtain the QP equation
\begin{align}
\left[\e_{\lm}-h^{(1)}_s(\br)\right]\phi_{\lm}(\br,s)
-\int\Sigma_s(\br,\br';\e_{\lm})\phi_{\lm}(\br',s)d\br' = 0,
\label{qpe'}
\end{align}
which is equivalent to Eqs. (\ref{QPeq1}) and (\ref{QPeq2}).
If the eigenvalues are $m$-fold degenerate, one can choose the corresponding $m$ QP wave functions
to be orthogonal to each other in this degenerate subspace.
Then, substituting Eq. (\ref{Lehman}) with $\om=\e_{\lm}$ into Eq. (\ref{Dyson_w}),
multiplying the resulting equation by the complex conjugate of one of the $m$ QP wave functions $\phi^\ast_{\lm}(\br',s)$,
and integrating with respect $\br'$, we derive Eq. (\ref{qpe'}).

With the aid of Eqs. (\ref{G1b}) and (\ref{phiTD}), the left-hand side (l.h.s.) of Eq. (\ref{S1}) can be written as
\begin{align}
& i\sum_{\mu}^{\rm occ}
\int\left(\int_{-\infty}^{t'} e^{-i\e_\mu t''} \Sigma_s(x,x'')\, dt''\right) \phi_{\mu}(\br'',s) d\br''
\, \phi^\ast_{\mu}(x',s)
%\theta(t'-t''),
\nonumber\\
- & i\sum_{\nu}^{\rm emp}
\int\left(\int_{t'}^{\infty} e^{-i\e_\nu t''} \Sigma_s(x,x'')\, dt''\right) \phi_{\nu}(\br'',s) d\br''
\, \phi^\ast_{\nu}(x',s),
%\theta(t''-t')
\label{Self1}
\end{align}
whereas the r.h.s. of Eq. (\ref{S1}) can be written as
\begin{align}
& i\theta(t'-t)\sum_{s''} \int h^{(2)}_{s''s}(\br'',\br)\bra{\Psi_{\g}^M}
\psi_s^{\dagger}(x')\psi_{s''}^{\dagger}(x'')\psi_{s''}(x'')\psi_s(x)
\ket{\Psi_{\g}^M}\bigr|_{t''=t}\, d\br''
\nonumber\\
- & i\theta(t-t')\sum_{s''} \int h^{(2)}_{s''s}(\br'',\br)\bra{\Psi_{\g}^M}
\psi_{s''}^{\dagger}(x'')\psi_{s''}(x'')\psi_s(x)\psi_s^{\dagger}(x')
\ket{\Psi_{\g}^M}\bigr|_{t''=t}\, d\br''
\nonumber\\
= & i\theta(t'-t)\sum_{\mu}^{\rm occ}\sum_{s''} \bra{\Psi_{\g}^M}\psi_s^{\dagger}(x')\ket{\Psi_{\mu}^{M-1}}
\int h^{(2)}_{s''s}(\br'',\br)\bra{\Psi_{\mu}^{M-1}}
\psi_{s''}^{\dagger}(x'')\psi_{s''}(x'')\psi_s(x)
\ket{\Psi_{\g}^M}\bigr|_{t''=t}\, d\br''
\nonumber\\
- & i\theta(t-t')\sum_{\nu}^{\rm emp}\sum_{s''} \int h^{(2)}_{s''s}(\br'',\br)\bra{\Psi_{\g}^M}
\psi_{s''}^{\dagger}(x'')\psi_{s''}(x'')\psi_s(x)
\ket{\Psi_{\nu}^{M+1}}\bigr|_{t''=t}\, d\br''
\, \bra{\Psi_{\nu}^{M+1}}\psi_s^{\dagger}(x')\ket{\Psi_{\g}^M}
\nonumber\\
= & i\theta(t'-t)\sum_{\mu}^{\rm occ}\sum_{s''}
\int h^{(2)}_{s''s}(\br'',\br)\bra{\Psi_{\mu}^{M-1}}
\psi_{s''}^{\dagger}(x'')\psi_{s''}(x'')\psi_s(x)
\ket{\Psi_{\g}^M}\bigr|_{t''=t}\, d\br''\, \phi^\ast_\mu(x',s)
\nonumber\\
- & i\theta(t-t')\sum_{\nu}^{\rm emp}\sum_{s''} \int h^{(2)}_{s''s}(\br'',\br)\bra{\Psi_{\g}^M}
\psi_{s''}^{\dagger}(x'')\psi_{s''}(x'')\psi_s(x)
\ket{\Psi_{\nu}^{M+1}}\bigr|_{t''=t}\, d\br''\, \phi^\ast_\nu(x',s).
\label{Self2}
\end{align}
Comparing these two equations, we find
\begin{subequations}
\begin{align}
& %\theta(t'-t) \sum_{\mu}^{\rm occ}
\int\left(\int_{-\infty}^{t'} e^{-i\e_\mu t''} \Sigma_s(x,x'')\, dt''\right) \phi_{\mu}(\br'',s) d\br''
%\, \phi^\ast_{\mu}(x',s)
- %\theta(t'-t)
f_\mu(x,s;t')
\nonumber\\
= & %\sum_{\mu}^{\rm occ}\sum_{s''}
\int h^{(2)}_{s''s}(\br'',\br)\bra{\Psi_{\mu}^{M-1}}
\psi_{s''}^{\dagger}(x'')\psi_{s''}(x'')\psi_s(x)
\ket{\Psi_{\g}^M}\bigr|_{t''=t}\, d\br''
%\, \phi^\ast_\mu(x',s)
\label{Self3a}
\end{align}
for $t'>t$ and
\begin{align}
& %\theta(t-t')
f_\nu(x,s;t')
- %\theta(t-t') \sum_{\nu}^{\rm emp}
\int\left(\int_{t'}^{\infty} e^{-i\e_\nu t''} \Sigma_s(x,x'')\, dt''\right) \phi_{\nu}(\br'',s) d\br''
%\, \phi^\ast_{\nu}(x',s)
\nonumber\\
= - & %\sum_{\nu}^{\rm emp}\sum_{s''}
\int h^{(2)}_{s''s}(\br'',\br)\bra{\Psi_{\g}^M}
\psi_{s''}^{\dagger}(x'')\psi_{s''}(x'')\psi_s(x)
\ket{\Psi_{\nu}^{M+1}}\bigr|_{t''=t}\, d\br''
%\, \phi^\ast_\nu(x',s).
\label{Self3b}
\end{align}
\label{Self3}
\end{subequations}
for $t>t'$. Here we put
\begin{subequations}
\begin{align}
%\theta(t'-t)
\sum_\mu^{\rm occ} f_\mu(x,s;t') \phi^\ast_{\mu}(x',s)
= %\theta(t'-t)
\sum_{\nu}^{\rm emp}
\int\left(\int_{t'}^{\infty} e^{-i\e_\nu t''} \Sigma_s(x,x'')\, dt''\right) \phi_{\nu}(\br'',s) d\br''
\, \phi^\ast_{\nu}(x',s)
\label{Self4a}
\end{align}
for $t'>t$ and
\begin{align}
%\theta(t-t')
\sum_\nu^{\rm emp} f_\nu(x,s;t') \phi^\ast_{\nu}(x',s)
= %\theta(t-t')
\sum_{\mu}^{\rm occ}
\int\left(\int_{-\infty}^{t'} e^{-i\e_\mu t''} \Sigma_s(x,x'')\, dt''\right) \phi_{\mu}(\br'',s) d\br''
\, \phi^\ast_{\mu}(x',s)
\label{Self4b}
\end{align}
\label{Self4}
\end{subequations}
for $t>t'$.
Note that $f_\mu(x,s;t')$ and $f_\nu(x,s;t')$ in Eq. (\ref{Self3}) are implicitly defined by Eq. (\ref{Self4}).
Since the r.h.s. of Eqs. (\ref{Self3a}) and (\ref{Self3b}) does not depend on $t'$,
the l.h.s. should not depend on $t'$ also.
Therefore, we can put $t'=\infty$ in Eqs. (\ref{Self3a}) and (\ref{Self4a}),
and $t'=-\infty$ in Eqs. (\ref{Self3b}) and (\ref{Self4b}).
Since $f_\mu(x,s;\infty)=f_\nu(x,s;-\infty)=0$, we finally obtain
\begin{subequations}
\begin{align}
& %\theta(t'-t) \sum_{\mu}^{\rm occ}
\int\left(\int_{-\infty}^{\infty} e^{-i\e_\mu t''} \Sigma_s(x,x'')\, dt''\right) \phi_{\mu}(\br'',s) d\br''
\nonumber\\
= & %\sum_{\mu}^{\rm occ}\sum_{s''}
\int h^{(2)}_{s''s}(\br'',\br)\bra{\Psi_{\mu}^{M-1}}
\psi_{s''}^{\dagger}(x'')\psi_{s''}(x'')\psi_s(x)
\ket{\Psi_{\g}^M}\bigr|_{t''=t}\, d\br''
%\, \phi^\ast_\mu(x',s)
\label{Self5a}
\end{align}
and
\begin{align}
& %\theta(t-t') \sum_{\nu}^{\rm emp}
\int\left(\int_{-\infty}^{\infty} e^{-i\e_\nu t''} \Sigma_s(x,x'')\, dt''\right) \phi_{\nu}(\br'',s) d\br''
%\, \phi^\ast_{\nu}(x',s)
\nonumber\\
= & %\sum_{\nu}^{\rm emp}\sum_{s''}
\int h^{(2)}_{s''s}(\br'',\br)\bra{\Psi_{\g}^M}
\psi_{s''}^{\dagger}(x'')\psi_{s''}(x'')\psi_s(x)
\ket{\Psi_{\nu}^{M+1}}\bigr|_{t''=t}\, d\br''.
%\, \phi^\ast_\nu(x',s).
\label{Self5b}
\end{align}
\label{Self5}
\end{subequations}
These equations are definitely identical to Eqs. (\ref{Sigma1a}) and (\ref{Sigma1b}).

\section{Perturbation Series for the Self-Energy} 
\label{Perturbation}

Next, we derive a concrete method to calculate the self-energy $\Sigma_s$.
Because here we discuss the energies at an instant time only,
even when the system is slowly varying in time, we simply drop all the time dependence
in the following expressions.

First, by the eqautions
\begin{subequations}
\begin{align}
& \psi_s(\br)\ket{\Psi_{\g}^M}
= \sum_\mu^{\rm occ} \ket{\Psi_{\mu}^{M-1}}\bra{\Psi_{\mu}^{M-1}}\psi_s(\br)\ket{\Psi_{\g}^M}
= \sum_\mu^{\rm occ} \ket{\Psi_{\mu}^{M-1}}\phi_\mu(\br,s)
= \sum_{\mu}^{\rm occ} a_{\mu}\ket{\Psi_{\g}^M}\phi_{\mu}(\br,s),
\label{psi=phi*a1}
\\
& \psi^\dagger_s(\br)\ket{\Psi_{\g}^M}
= \sum_\nu^{\rm emp} \ket{\Psi_{\nu}^{M+1}}\bra{\Psi_{\nu}^{M+1}}\psi^\dagger_s(\br)\ket{\Psi_{\g}^M}
= \sum_\nu^{\rm emp} \ket{\Psi_{\nu}^{M+1}}\phi^\ast_\nu(\br,s)
= \sum_{\nu}^{\rm emp} a^\dagger_{\nu}\ket{\Psi_{\g}^M}\phi^\ast_{\nu}(\br,s),
\label{psi=phi*a2}
\end{align}
\end{subequations}
and $a_{\nu}\ket{\Psi_{\g}^M} = a^\dagger_{\mu}\ket{\Psi_{\g}^M} = 0$
for an empty QP state $\nu$ and an occupied QP state $\mu$,
we introduce the annihilation and creation operators, $a_\lm$ and $a^\dagger_\lm$, as
\begin{subequations}
\begin{align}
\psi_{s} (\br) & =  \sum_\lambda \phi_\lambda (\br,s) a_\lambda,
\label{psi=phi*a1}
\\
\psi^\dagger_{s'} (\br') & =  \sum_{\lambda'} \phi^\ast_{\lambda'}(\br',s') a^\dagger_{\lambda'},
\label{psi=phi*a2}
\end{align}
\end{subequations}
where $\lambda$, $\lambda'$ includes both $\mu$ and $\nu$. 
Then, the anticommutation relations of the field operators are written as
\begin{subequations}
\begin{align}
\left\{ \psi_s(\br), \psi^\dagger_{s'}(\br') \right\} & =
\sum_{\lambda\lambda'} \phi_\lambda(\br,s) \phi^\ast_{\lambda'}(\br',s')
\left \{ a_\lambda, a^\dagger_{\lambda'} \right\} = \delta(\br - \br')\delta_{ss'},
\\
\left\{ \psi_s (\br), \psi_{s'}(\br') \right\} & =
\sum_{\lambda\lambda'} \phi_\lambda(\br,s) \phi_{\lambda'}(\br',s')
\left \{ a_\lambda, a_{\lambda'} \right\} = 0,
\\
\left\{ \psi^\dagger_s(\br), \psi^\dagger_{s'}(\br') \right\} & =
\sum_{\lambda\lambda'} \phi^\ast_\lambda(\br,s) \phi^\ast_{\lambda'}(\br',s')
\left \{ a^\dagger_\lambda, a^\dagger_{\lambda'}  \right\} = 0.
\end{align}
\label{commutation_psi}
\end{subequations}
From these equations, we can assume that the QP annihilation and creation operators satisfy the anticommutation relations
\begin{align}
\left\{  a_\lambda , a^\dagger_{\lambda'}  \right\}  = \delta_{\lambda \lambda'}, \ \
\left\{  a_\lambda , a_{\lambda'}  \right\}  = 0, \ \
\left\{  a^\dagger_\lambda , a^\dagger_{\lambda'}  \right\}  = 0.
\label{commutation_a}
\end{align}
Note that these relations hold even when
the QP states $\lm, \lm'$ are not orthogonal to each other,
because Eq. (\ref{commutation_psi}) with Eq. (\ref{commutation_a}) satisfies
the completeness condition (\ref{complete}) only.

\begin{figure}[hbtp]
\begin{center}
\includegraphics[width=120mm]{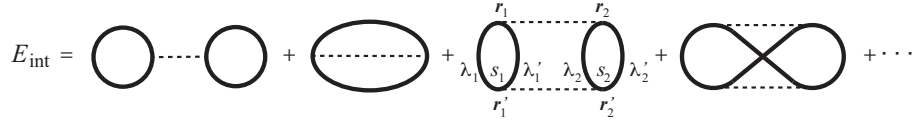}
\caption{Diagrammatic representation of the interaction energy $E_{\rm int}$.}
\label{fig:Eint}
\end{center}
\end{figure}

Second, we introduce a hypothetical non-interacting system, where the QPs are not interacting each other,
although in real systems the QPs generally interact with each other at least at short distances.
This hypothetical system is described by the Hamiltonian
\begin{align}
H' = H^{(1)} + H^{(2)}_{\rm self}
\label{H_hypothetical}
\end{align}
where $H^{(2)}_{\rm self}$ is given by (\ref{H_2body}) whose operation is restricted to construct
the self-energy $\Sigma_s$ only and ineffective for any other interaction.
Its eigenvalue equation is given by
\begin{align}
H' \ket{\Phi^{M+n}_\alpha} = {E'}^{M+n}_\alpha \ket{\Phi^{M+n}_\alpha}
\label{eigenvalue_Phi}
\end{align}
where $M + n$ indicates the total number of electrons in this system.
When $n=0$ and $\alpha=\g$, we assume that $\ket{\Phi_{\g}^M}$ has the same energy eigenvalue
${E'}^M_{\g}=E_{\g}^M$ as the true system. 
All the other states of the hypothetical system are constructed
by operating several $a_{\lm}$'s and $a^{\dagger}_{\lm}$'s on $\ket{\Phi_{\g}^M}$.
We assume $H'$ as the unperturbed Hamiltonian and 
\begin{align}
H'' = H^{(2)} - H^{(2)}_{\rm self}
\label{H_perturbation}
\end{align}
as a perturbation, and adopt Brillouin--Wigner's perturbation theory,
which is well known to be applicable not only to the ground state
but also to any excited eigenstate of the full Hamiltonian.
Multiplying both sides of
\begin{align}
\left(E_{\g}^M - H'\right)\ket{\Psi_{\g}^M} = H''\ket{\Psi_{\g}^M},
\end{align}
by the projection operator
\begin{align}
P = 1 - \ket{\Phi_{\g}^M}\bra{\Phi_{\g}^M},
\end{align}
and noting that $H'$ commutes with $P$, we have
\begin{align}
\left(E_{\g}^M - H'\right)P\ket{\Psi_{\g}^M} = PH''\ket{\Psi_{\g}^M}.
\label{mid1}
\end{align}
Equation (\ref{mid1}) can be rewritten as
\begin{align}
\ket{\Psi_{\g}^M} = \ket{\Phi_{\g}^M}\braket{\Phi_{\g}^M}{\Psi_{\g}^M} + \frac{1}{E_{\g}^M - H'}PH''\ket{\Psi_{\g}^M}.
\end{align}
Then, by iteration, we have
\begin{align}
\frac{\ket{\Psi_{\g}^M}}{\braket{\Phi_{\g}^M}{\Psi_{\g}^M}}
 & = \ket{\Phi_{\g}^M} + \frac{1}{E_{\g}^M - H'}PH''\ket{\Phi_{\g}^M}
 \nonumber\\
 & + \frac{1}{E_{\g}^M - H'}PH''\frac{1}{E_{\g}^M - H'}PH''\ket{\Phi_{\g}^M} + \cdots.
\label{Brillouin1}
\end{align}
So, the expectation value of $A(\br,\br')=\psi_s^\dagger(\br)\psi_{s'}^\dagger(\br')\psi_{s'}(\br') \psi_s(\br)$
can be evaluated as
\begin{align}
\langle A(\br,\br')\rangle
 & = \bra{\Phi_{\g}^M}A(\br,\br')\ket{\Phi_{\g}^M}
 + \bra{\Phi_{\g}^M}A(\br,\br')\frac{1}{E_{\g}^M - H'}PH^{(2)}\ket{\Phi_{\g}^M}
 \nonumber\\
 & + \bra{\Phi_{\g}^M}A(\br,\br')\frac{1}{E_{\g}^M - H'}PH^{(2)}\frac{1}{E_{\g}^M - H'}PH^{(2)}
 \ket{\Phi_{\g}^M} + \cdots,
 \label{Brillouin2}
\end{align}
where we replaced $H''$ with $H^{(2)}$ without restriction because it is obvious that
these interactions are not related to the construction of the self-energy but simply
related to the construction of the expectation value of $A(\br,\br')$.
Then, multiplying the both sides of Eq. (\ref{Brillouin2}) by $\frac{1}{2}h_{ss'}^{(2)}(\br,\br')$,
summing over $s,s'$, and integrating with respect to $\br$ and $\br'$, we have
\begin{align}
E_{\rm int}
 & = \bra{\Phi_{\g}^M}H^{(2)}\ket{\Phi_{\g}^M}
 + \bra{\Phi_{\g}^M}H^{(2)}\frac{1}{E_{\g}^M - H'}PH^{(2)}\ket{\Phi_{\g}^M}
 \nonumber\\
 & + \bra{\Phi_{\g}^M}H^{(2)}\frac{1}{E_{\g}^M - H'}PH^{(2)}\frac{1}{E_{\g}^M - H'}PH^{(2)}
 \ket{\Phi_{\g}^M} + \cdots.
 \label{Brillouin3}
\end{align}
In the $k$-th order term in this perturbation series, there are $4(k+1)$ field operators
sandwiched by $\bra{\Phi_{\g}^M}$ and $\ket{\Phi_{\g}^M}$, which can be expanded as 
(\ref{psi=phi*a1}) and (\ref{psi=phi*a2}).
For example, the first term in the r.h.s. of Eq. (\ref{Brillouin3}) is expressed as
\begin{align}
\frac{1}{2}\sum_{ss'}\sum_{\mu_1\mu_2\mu_3\mu_4}^{\rm occ}\int h_{ss'}^{(2)}(\br,\br')
\phi^\ast_{\mu_1}(\br,s)\phi^\ast_{\mu_2}(\br',s')\phi_{\mu_3}(\br',s')\phi_{\mu_4}(\br,s)d\br d\br'
\bra{\Phi_{\g}^M}a^{\dagger}_{\mu_1}a^{\dagger}_{\mu_2}a_{\mu_3}a_{\mu_4}\ket{\Phi_{\g}^M}.
\label{1st}
\end{align}
In this matrix element, the QP states $\mu_4$ and $\mu_3$ are removed from
and $\mu_2$ and $\mu_1$ are added to the $M$-particle state $\ket{\Phi_{\g}^M}$,
and after these operations,
the $M$-particle state must be restored to the original state $\ket{\Phi_{\g}^M}$;
otherwise this matrix element becomes zero.
That is, either ($\mu_1=\mu_4$ and $\mu_2=\mu_3$) or ($\mu_1=\mu_3$ and $\mu_2=\mu_4$)
should be satisfied.
Of course, this first order approximation rigorously corresponds to the Hartree--Fock energy
in the extended HFA \cite{Morokuma}.
In fact, Eq. (\ref{1st}) becomes
\begin{align}
E_{\rm int}^{\rm HF} & =
\frac{1}{2}\sum_{ss'}\sum_{\mu\mu'}^{\rm occ}\int h_{ss'}^{(2)}(\br,\br')
\left|\phi_{\mu}(\br,s)\right|^2 \left|\phi_{\mu'}(\br',s')\right|^2 d\br d\br'
\\
& - \frac{1}{2}\sum_{s}\sum_{\mu\mu'}^{\rm occ}\int h_{ss}^{(2)}(\br,\br')
\phi^\ast_{\mu}(\br,s)\phi^\ast_{\mu'}(\br',s)\phi_{\mu}(\br',s)\phi_{\mu'}(\br,s)d\br d\br'.
\label{HF}
\end{align}
Its analogy holds for all orders.
That is, all the intermediate QP states $\lm_1, \lm_2, \lm_3, \lm_4, \cdots$ must be pairwise
($\lm_i$ can be either occupied states $\mu_i$ or empty states $\nu_i$);
each pair consists of one annihilation operator $a_{\lm_i}$
and one creation operator $a^{\dagger}_{\lm_i}$ of the same QP state $\lm_i$.
All possible combinations of the product of the pairs have to be summed up
to evaluate the total contribution at this order. 
Because of the existence of the projection operator $P$ in (\ref{Brillouin3}),
the intermediate states cannot be the initial state $\ket{\Phi_{\g}^M}$.
Moreover, the energy denominator in (\ref{Brillouin3}) is just a simple
addition or subtraction of the QP energies $\e_{\lm}$'s.
If we draw a pair of the QP state by a thick sold line and $H^{(2)}$ by a dotted line,
(\ref{Brillouin3}) is expressed by the diagrams shown in Fig. \ref{fig:Eint}.
This expression is exactly the same as the interaction energy expressed by the skeleton diagrams,
i.e., the diagrams expressed by the clothed Green's function (a thick solid line in Fig. \ref{fig:Eint}),
in the usual MBPT \cite{Feynman,Luttinger,Pines,Nozieres}. % \cite{Feynman,Fetter,Luttinger,GellMann,NozieresPines,Nozieres}.
They also correspond to the same order contribution
in the M{\o}ller--Plesset theory if the initial excited-state configuration
and the full self-consistency are imposed in the latter.
Thus we derived exactly the same formula only by using the true eigenstates
of the full Hamiltonian and their derivatives such as the QP energies and QP wave functions.
An important point here is that we have never assumed the orthogonality of the QP wave functions
in these equations.

\begin{figure}[hbtp]
\begin{center}
\includegraphics[width=100mm]{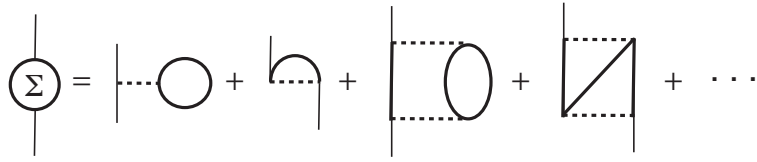}
\caption{Diagrammatic representation of the self-energy $\Sigma_s$.
Two external thin solid lines are additionally drawn to indicate the dependence of the self-energy
on the two positions $(\br,s)$ and $(\br',s)$.}
\label{fig:SelfEnergy}
\end{center}
\end{figure}

Third, in order to obtain the expression for the self-energy $\Sigma_s$,
we compare the expression (\ref{Brillouin3}) with the Galitskii and Migdal's formula of the interaction energy (\ref{Eint}).
Looking at Eq. (\ref{Eint}), we find that if we drop $\phi_{\mu}(\br,s)\phi^\ast_{\mu}(\br',s)$ from (\ref{Brillouin3}),
{\it i.e.}, a thick solid line from Fig. \ref{fig:Eint},
we should have the self-energy $\Sigma_s(\br,\br';\e_{\mu})$ equivalent to Fig. \ref{fig:SelfEnergy}.
In Fig. \ref{fig:SelfEnergy}, we drew two external thin solid lines to indicate the dependence of the self-energy $\Sigma_s$
on the two positions $(\br,s)$ and $(\br',s)$.
This is again exactly the same skeleton diagrams for the self-energy in the MBPT \cite{Pines,Nozieres}.
When we remove a pair of $\phi_{\mu}(\br,s)\phi^*_{\mu}(\br',s)$ from the product of $H^{(2)}$'s
in the expression of $E_{\rm int}$ (\ref{Brillouin3}),
we also have to remove the summation with respect to this special QP state, $\mu$.
Therefore, the corresponding QP energy $\e_{\mu}$ appearing in some
parts of the energy denominators $(E^M_{\g} - H')^{-1}$ remains unsummed
with the special index $\mu$ after taking the summations with respect to all the intermediate QP states,
resulting in the $\e_{\mu}$ dependence in the final expression.
Consequently, the resulting self-energy must have the explicit
$\e_{\mu}$ dependence, where $\mu$ corresponds to the QP state of the external thin line in Fig. \ref{fig:SelfEnergy}.

Last, we make a brief comment on the relationship between the present formulation
using the resolvent operator $(E^M_{\g} - H')^{-1}$ and the formulation using the clothed Green's function (\ref{Lehman}).
To see this relationship, we insert the completeness condition
\begin{align}
\sum_{\alpha}\ket{\Phi^M_{\alpha}}\bra{\Phi^M_{\alpha}}=1
\end{align}
in between each two $H^{(2)}$'s in Eq. (\ref{Brillouin3}).
Then $H^{(2)}$ is replaced by the sum over all possible intermediate QP states
$\lm_1, \lm_2, \lm_3, \lm_4$ of the product of four QP wave functions
and the matrix element $\bra{\Phi_{\lm}^M}\cdots\ket{\Phi_{\lm}^M}$
replaced with $\bra{\Phi_{\alpha}^M}\cdots\ket{\Phi_{\beta}^M}$ in a form (\ref{1st}).
(Note that the number of electrons remains $M$ in the intermediates states $\alpha$ and $\beta$,
because $H^{(2)}$ has two annihilation and two creation operators.)
Then, all the QP states $\lm_i$ are made pairwise, each of which has a form $\phi_{\lm}(\br,s)\phi^*_{\lm}(\br',s)$
identical to the numerator of the Green's function (\ref{Lehman}).
Simultaneously, each resolvent operator is replaced with its expectation value $(E^M_{\g} - E'^M_{\alpha})^{-1}$
sandwiched by an intermediate state $\ket{\Phi^M_{\alpha}}$,
which corresponds to the denominator of the result of the $\om$-integration of the even-number product of the Green's functions.
As an example, let us consider the third term of Fig. \ref{fig:Eint}.
It consists of two bubble diagrams, each of which corresponds to the polarization function $P(\br_i,\br'_i;\om_i)$
($i=1,2$) in the random-phase approximation (RPA), connected by two dotted lines,
$h_{s_1s_2}^{(2)}(\br_1,\br_2)$ and $h_{s_1s_2}^{(2)}(\br'_1,\br'_2)$.
The polarization function is evaluated as
\begin{align}
%P(\br,\br';\om) &= -i\hbar\sum_s\int_{-\infty}^{\infty}G_s(\br,\br';\om+\om')G_s(\br',\br;\om')\frac{d\om'}{2\pi}
P(\br,\br';\om) &= -i\sum_s\int_{-\infty}^{\infty}G_s(\br,\br';\om+\om')G_s(\br',\br;\om')\frac{d\om'}{2\pi}
\nonumber\\
& = \sum_{\lm\lm'}\frac{\phi^*_{\lm}(\br,s)\phi_{\lm'}(\br,s)\phi^*_{\lm'}(\br',s)\phi_{\lm}(\br',s)}
%{\hbar\om - \e_{\lm} + \e_{\lm'} - i\delta_{\lm}}[f(\e_{\lm'})-f(\e_{\lm})]
{\om - \e_{\lm} + \e_{\lm'} - i\delta_{\lm}}[f(\e_{\lm'})-f(\e_{\lm})]
\label{P}
\end{align}
as usual, where $f(\e)$ is the Fermi distribution function.
Multiplying $P(\br_1,\br'_1;\om)$ by $P(\br'_2,\br_2;\om)=P(\br_2,\br'_2;-\om)$ and $i/2\pi$,
and integrating it with respect to $\om$ from $-\infty$ to $\infty$,
we obtain a possible term,
\begin{align}
\frac{
\bra{\Phi_{\g}^M}a^{\dagger}_{\lm_1}a^{\dagger}_{\lm_2}a_{\lm'_2}a_{\lm'_1}\ket{\Phi_{\alpha}^M}
\bra{\Phi_{\alpha}^M}a^{\dagger}_{\lm'_1}a^{\dagger}_{\lm'_2}a_{\lm_2}a_{\lm_1}\ket{\Phi_{\g}^M}
}{E^M_{\g} - E'^M_{\alpha}}
= \frac{\delta_{\lm_1}^{\rm occ}\delta_{\lm'_1}^{\rm emp}\delta_{\lm_2}^{\rm occ}\delta_{\lm'_2}^{\rm emp}}
{\e_{\lm_1}-\e_{\lm'_1}+\e_{\lm_2}-\e_{\lm'_2}} %[f(\e_{\lm'_1})-f(\e_{\lm_1})][f(\e_{\lm'_2})-f(\e_{\lm_2})]
\label{den1}
\end{align}
whose denominator is identical to the expectation values of the resolvent operator $(E^M_{\g} - H')^{-1}$
in the second term of Eq. (\ref{Brillouin3}). Final result can be obtained by operating
\begin{align}
& \frac{1}{4}
\sum_{\lm_1}\sum_{\lm'_1}\sum_{\lm_2}\sum_{\lm'_2}
\sum_{s_1}\sum_{s_2}\int d\br_1 d\br_2 d\br'_1 d\br'_2 h_{s_1s_2}^{(2)}(\br_1,\br_2) h_{s_1s_2}^{(2)}(\br'_1,\br'_2)
\nonumber\\
&
\times
\phi^*_{\lm_1}(\br_1,s_1)\phi^*_{\lm_2}(\br_2,s_2)\phi_{\lm'_2}(\br_2,s_2)\phi_{\lm'_1}(\br_1,s_1)\times
\phi^*_{\lm'_1}(\br'_1,s_1)\phi^*_{\lm'_2}(\br'_2,s_2)\phi_{\lm_2}(\br'_2,s_2)\phi_{\lm_1}(\br'_1,s_1).
\label{phi*8}
\end{align}
The eight product of the QP wave functions comes from the expansions (\ref{psi=phi*a1}) and (\ref{psi=phi*a2}).
All these operations correspond to taking the trace, $\Tr[Ph^{(2)}Ph^{(2)}]$, in the Green's function method \cite{Kuwahara}.

%\vskip3mm

\section{Test Calculations}
\label{test1}

In order to verify the applicability of the QP theory for some initial excited eigenstates,
let us first consider a hydrogen atom with $N=1$
and for the case $M=0$ as the simplest case.
Its ($M=0$)-electron excited state is none other than a proton.
Then, the lowest-energy (empty) QP state having $M+1=1$ electron
is the hydrogen 1s orbital.
The self-consistent calculation
surely gives this result exactly,
because there is no occupied electron.
Note that there is no Hartree term, no exchange term, and no higher-order terms.
This is so even in the LDA or the generalized gradient approximation (GGA) of DFT.
The same thing holds not only for the 1s state but also for any excited eigenstates
of the hydrogen atom.
Similar situation occurs also for an alkali metal atom like lithium,
sodium, and potassium, if we can ignore the core contributions.
In this sense, the present excited-state formalism has
a chance to serve
an extremely good result compared to the ordinary ground-state formalism
in some special cases.

%As a more general case, let us
Next we consider a general isolated atom A or molecule M.
It is obvious that the ionization potential (IP) of A or M is identical to
the electron affinity (EA) of its cation A$^+$ or M$^+$,
provided that the atomic geometries of M and M$^+$ are the same
(This is not a bad approximation for typical molecules).
Similarly, there are several other identities between the electron removal
%energy 
and 
%the energy gain at electron
absorption
energies
of cationic, anionic
and photoabsorbed states of A or M.
These identities allow us to check the validity of the present theory.
%Note that isolated
%Isolated
%atoms are good target for this check because there is no ambiguity on atomic geometry.
%Highly excited atoms and ions have been a subject in atomic physics \cite{LebedovBeigman}
%and has been investigated by using the orthogonality-constrained approach \cite{Blushkov},
%which must be performed one-by-one for constructing an excited state orthogonal to the ground state.
%Other approaches such as the equation of motion (EOM) approach \cite{Bartlett,WangTuWang}
%or other approach \cite{NicolaidesBeck} can handle, e.g., one photon-absorbed states
%as well as the $GW$ + Bethe--Salpeter approach \cite{HybertsenLouie,RohlfingLouie,Ohno2} in the MBPT. 
Here, we perform the perturbation expansion starting from the cationic, anionic,
or photoabsorbed state as well as the neutral ground state
of %an isolated atom
A or M to calculate the EQP energies,
which should coincide with each other. % if they correspond to the same physical energy.

\begin{widetext}

\begin{table}[hptb]
\begin{center}
\caption{$G_0W_0$ QP energies of neutral (A), cationic (A$^+$),
anionic (A$^-$), and photoabsorbed (A$^*$) states of lithium, aluminum, and beryllium atoms in units of eV.
The right atom/ion has one more electron than the left atom/ion; and
the left and the right have $N(-1)$ and $N(+1)$ electron configurations, respectively
(${}^{2S+1}L$ states are also listed);
In the same row, the QP energies of the forward, $\rightarrow$, electron absorption process 
and the reverse, $\leftarrow$, electron release process should coincide with each other.
%and also with the reference values \cite{Ref5,Ref6,Ref7,Ref8,Ref9,%Ref10,
%Ref11,Ref12}.
The values inside the parentheses indicate the usual $G_0W_0$ QP energy calculated
for the neutral atom at the ground state. EC in the superscript in the Reference values
means the value calculated by the EOM coupled cluster singles and doubles (EOM-CCSD)
with aug-cc-pV5Z basis set using Gaussian 09 \cite{Gaussian}.
All the other reference values$^{a-g}$ are experimental data \cite{Ref5,Ref6,Ref7,Ref8,Ref9,%Ref10,
Ref11}.}
\begin{threeparttable}
{\tabcolsep = 3mm
\begin{tabular}{lccclccrrr}\hline\hline
%--------------------------------------------------------------------------------
Atom & $N(-1)$ Electron & ${}^{2S+1}L$ & & Atom & $N(+1)$ Electron & ${}^{2S+1}L$ & \multicolumn{3}{c}{QP energies (eV)} \\ \cline{8-10}
/ Ion &Configuration & State & & / Ion &Configuration& State & $\rightarrow${ } & $\leftarrow${ } { } & Reference \\ \hline
%--------------------------------------------------------------------------------
{ } Li$^+$    & $(1s)^2$     & ${}^1$S & \lra & { } Li        & $(1s)^2(2s)$     & ${}^2$S &  5.1 &  (5.6) &  5.39\tnote{a}{ }{ } \\ \hline
%--------------------------------------------------------------------------------
{ } Li$^+$    & $(1s)^2$     & ${}^1$S & \lra & { } Li$^*$    & $(1s)^2(2p)$     & ${}^2$P &  3.0 &  3.2 &  3.54\tnote{b}{ }{ } \\ \hline
%--------------------------------------------------------------------------------
{ } Li$^{+*}$ & $(1s)(2s)$   & ${}^1$S & \lra & { } Li        & $(1s)^2(2s)$     & ${}^2$S & 63.9 & (63.4) & 64.46\tnote{b}{ }{ } \\ \hline
%--------------------------------------------------------------------------------
{ } Li$^{+*}$ & $(1s)(2p)$   & ${}^1$P & \lra & { } Li$^*$    & $(1s)^2(2p)$     & ${}^2$P & 65.0 & 66.0 & 65.76\tnote{b}{ }{ } \\ \hline
%--------------------------------------------------------------------------------
{ } Li        & $(1s)^2(2s)$ & ${}^2$S & \lra & { } Li$^-$    & $(1s)^2(2s)^2$   & ${}^1$S &  (0.6) &  0.3 &  0.61\tnote{c}{ }{ } \\ \hline
%--------------------------------------------------------------------------------
{ } Al$^+$    & $(3s)^2$     & ${}^1$S & \lra & { } Al        & $(3s)^2(3p)$     & ${}^2$S &  5.7 &  (5.6) &  5.98\tnote{d}{ }{ } \\ \hline
%--------------------------------------------------------------------------------
{ } Al$^+$    & $(3s)^2$     & ${}^1$S & \lra & { } Al$^*$    & $(3s)^2(4s)$     & ${}^2$P &  2.6 &  3.1 &  2.84\tnote{e}{ }{ } \\ \hline
%--------------------------------------------------------------------------------
{ } Al$^{+*}$ & $(3s)(3p)$   & ${}^1$S & \lra & { } Al		  & $(3s)^2(3p)$     & ${}^2$P & 12.6 & (10.2) & 13.40\tnote{e}{ }{ } \\ \hline
%--------------------------------------------------------------------------------
{ } Al        & $(3s)^2(3p)$ & ${}^2$P & \lra & { } Al$^-$    & $(3s)^2(3p)^2$   & ${}^3$S &  (0.3) & -{ } { } &  0.43\tnote{f}{ }{ } \\ \hline
%--------------------------------------------------------------------------------
{ } Al        & $(3s)^2(3p)$ & ${}^2$P & \lra & { } Al$^{-*}$ & $(3s)^2(3p)(4s)$ & ${}^3$P & (-0.1) &  0.5 &  0.2$^{\rm EC}$  \\ \hline
%--------------------------------------------------------------------------------
{ } Al$^*$    & $(3s)^2(4s)$ & ${}^2$S & \lra & { } Al$^{-*}$ & $(3s)^2(3p)(4s)$ & ${}^3$P &  3.5 &  2.7 &  3.3$^{\rm EC}$ \\ \hline
%--------------------------------------------------------------------------------
%{ } Be$^+$    & $(2s)$       & ${}^2$S & \lra & { } Be        & $(2s)^2$         & ${}^1$S &  9.0 &  (9.3) &  9.32\tnote{g}{ }{ } \\ \hline
{ } Be$^+$    & $(2s)$       & ${}^2$S & \lra & { } Be        & $(2s)^2$         & ${}^1$S &  9.3 &  (9.3) &  9.32\tnote{f}{ }{ } \\ \hline
%--------------------------------------------------------------------------------
%Be$^+$    & $(2s)$       & ${}^2$S & \lra & { } Be$^*$    & $(2s)(2p)$       & ${}^1$P &  4.7 &  5.4 &  4.40\tnote{h}{ }{ } \\ \hline
{ } Be$^+$    & $(2s)$       & ${}^2$S & \lra & { } Be$^*$    & $(2s)(2p)$       & ${}^1$P &  4.2 &  5.4 &  4.40\tnote{g}{ }{ } \\ \hline
%--------------------------------------------------------------------------------
{ } Be$^+$    & $(2s)$       & ${}^2$S & \lra & { } Be$^*$    & $(2s)(2p)$       & ${}^3$P &  6.4 &  6.2 &  6.59\tnote{g}{ }{ } \\ \hline
%--------------------------------------------------------------------------------
\hline
\end{tabular}
}
\begin{tablenotes}\scriptsize
\item[a] Reference \cite{Ref5}.
\item[b] References \cite{Ref5,Ref6}.
\item[c] References \cite{Ref7}.
\item[d] Reference \cite{Ref8}.
\item[e] References \cite{Ref8,Ref9}.
\item[f] Reference \cite{Ref11}.
\item[h] References %\cite{Ref11,Ref12}.
\cite{Ref6,Ref11}.
\end{tablenotes}
\end{threeparttable}
\label{Table:atoms}
\end{center}
\end{table}

\end{widetext}

\begin{table}[htbp]
\begin{center}
\caption{$G_0W_0$ QP energies of neutral (M), cationic (M$^+$),
and photoabsorbed (M$^*$) states of nitrogen, oxygen, and lithium molecules in units of eV.
The right molecule has one more electron than the left molecule; and
the left and the right have $N-1$ and $N$ electron configurations, respectively (electronic states are also listed);
In the same row, the QP energies of the forward, $\rightarrow$, electron absorption process 
and the reverse, $\leftarrow$, electron release process should coincide with each other and
also with the reference values \cite{Molecule1,Molecule2,Molecule3,Molecule4}.
The values inside the parentheses indicate the usual $G_0W_0$ QP energy calculated
for the neutral atom at the ground state.}
\begin{threeparttable}
{%\tabcolsep = 3mm
\begin{tabular}{cccccrrr}\hline\hline
%--------------------------------------------------------------------------------
Molecule & State & & Molecule & State & \multicolumn{3}{c}{QP energies (eV)} \\ \cline{6-8}
         &       & &          &       & $\rightarrow${ } & $\leftarrow${ } { } & Reference \\ \hline
%--------------------------------------------------------------------------------
{ } N$_2^+$	& ${}^2\Sigma^+_g$	& \hs\lra	& { } N$_2$		& ${}^1\Sigma_g^+$	& 15.69 & (15.40)& 15.58\tnote{a}{ }{ } \\ \hline
%--------------------------------------------------------------------------------
{ } N$_2^+$	& ${}^2\Sigma_g^+$	& \hs\lra	& { } N$_2^*$	& ${}^1\Pi_g$		&  7.28 &   7.21 &  6.99\tnote{b}{ }{ } \\ \hline
%--------------------------------------------------------------------------------
{ } O$_2^+$	& ${}^2\Pi_g$		& \hs\lra	& { } O$_2$		& ${}^3\Sigma_g^-$	& 12.16 & (12.35)& 12.30\tnote{c}{ }{ } \\ \hline
%--------------------------------------------------------------------------------
{ } O$_2^+$	& ${}^2\Pi_g$		& \hs\lra	& { } O$_2^*$	& ${}^3\Pi_g$		&  3.80 &   3.73 &  4.17\tnote{d}{ }{ } \\ \hline
%--------------------------------------------------------------------------------
{ } Li$_2^+$& ${}^2\Sigma_g^+$	& \hs\lra	& { } Li$_2$	& ${}^1\Sigma_g^-$	&  5.30 &  (5.32)&  5.11\tnote{e}{ }{ } \\ \hline
%--------------------------------------------------------------------------------
{ } Li$_2^+$& ${}^2\Sigma_g^+$	& \hs\lra	& { } Li$_2^*$	& ${}^1\Sigma_u^-$	&  3.74 &   3.95 &  3.35\tnote{f}{ }{ } \\ \hline
%--------------------------------------------------------------------------------
\hline
\end{tabular}
}
\begin{tablenotes}\scriptsize
\item[a] Reference \cite{Molecule1}.
\item[b] References \cite{Molecule1,Molecule2}.
\item[c] Reference \cite{Molecule3}.
\item[d] References \cite{Molecule2,Molecule3}.
\item[e] Reference \cite{Molecule4}.
\item[f] References \cite{Molecule2,Molecule4}.
\end{tablenotes}
\end{threeparttable}
\label{Table:molecules}
\end{center}
\end{table}

\begin{figure}[hbtp]
\begin{center}
\includegraphics[width=59mm]{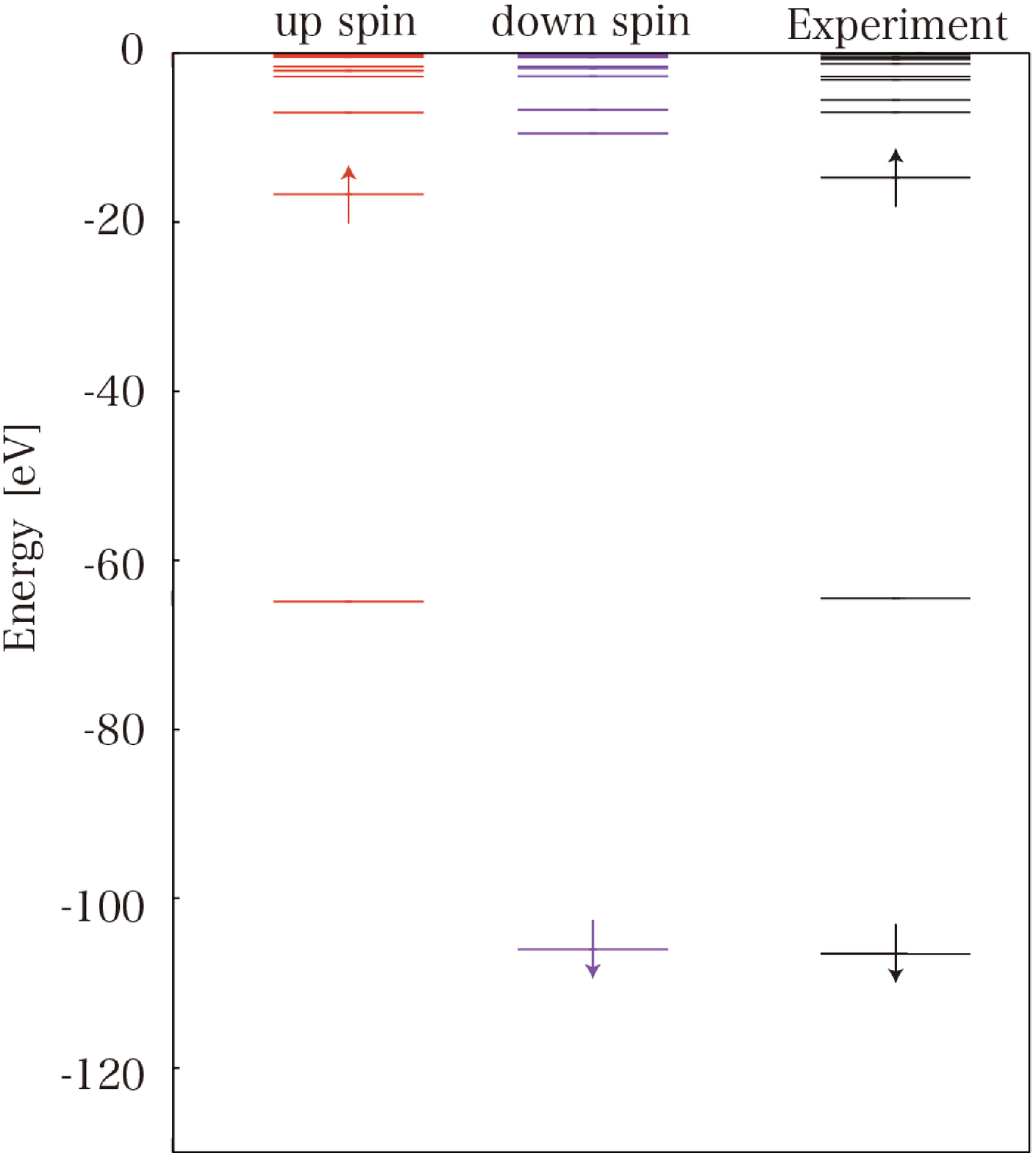}
\caption{QP energy spectra of Li$^{+*}$
calculated by $G_0W_0$ in units of eV.
Experimental data are taken from Refs. \cite{Ref5,Ref6,Ref13,Ref14}.
Arrows represent electrons (spin) occupying the level.}
\label{Fig1}
\end{center}
\end{figure}

We use the standard one-shot $GW$ approximation ($G_0W_0$) by beginning with the LDA \cite{HybertsenLouie}
for both the ground-state and excited-state configurations of isolated lithium, aluminum, and beryllium atoms
and nitrogen, oxygen and lithium (diatomic) molecules with the bond lengths fixed at 1.106~\AA,
1.216~\AA, and 2.723~\AA, respectively, for N$_2$, O$_2$, and Li$_2$ (calculated with B3LYP/6-31G* \cite{NIST}).
We use the all-electron mixd basis, TOMBO code \cite{TOMBO}, %\cite{Ohno2,TOMBO},
in which both plane waves (PWs) and atomic orbitals (AOs) are used as basis functions.
%Unit cell is fcc with an edge length of 18~\AA{ }for Li, Li$_2$, and Al, 12~\AA{ }for
The face-centered cubic (fcc) unit cell with an edge length of 18~\AA{ }is used for Li, Li$_2$, and Al,
and that of 12~\AA{ }is used for
Be, N$_2$, and O$_2$.
Cutoff energies for PWs, the Fock exchange and the correlation part
of the self-energy (as well as the polarization part) are set
as 11.05~Ry, 44.22~Ry, 11.05~Ry for Li, Li$_2$, and Al, 19.65~Ry, 78.61~Ry, 19.65~Ry for Be and N$_2$,
and 27.71~Ry, 122.83~Ry, 30.71~Ry for O$_2$, respectively.
In the summation in the correlation part, 400 levels are included for Li, Al, and Be;
3000, 6000, and 7000 levels are included, respectively, for Li$_2$, N$_2$, and O$_2$.
We use the generalized plasmon pole model \cite{HybertsenLouie} to avoid the $\omega$ integration.
The energy dependence of the self-energy is dealt by introducing the $Z$ factor
(linearization around the Kohn-Sham eigenvalues) as usual.

The computational time scales as $O(N_{\rm emp}N_{\rm occ}N_G^2)$
where $N_{\rm emp}$, $N_{\rm occ}$, and $N_G$ are the numbers of empty states,
occupied states, and $\G$ vectors needed for the non-local Fourier components of
the polarization function and the correlation part of the self-energy.

There is no essential difference between the QP theories using the neutral ground state
and the other (neutral or charged) excited states as the initial state,
and the GW calculations for the latter excited-state case can be simply performed
by just changing the order of levels after the diagonalization.
However, in order to compare the results, we will call the former neutral ground-state case
the ordinary QP (OQP) theory and the latter excited-state case the extended QP (EQP) theory.

The calculated results of the QP energies of several QP states
are listed in Table \ref{Table:atoms} for Li, Al, and Be, and in Table \ref{Table:molecules} for N$_2$, O$_2$, and Li$_2$,
together with the reference values.
EC in Table~\ref{Table:atoms} means the value calculated by the EOM coupled cluster singles and doubles (EOM-CCSD)
with aug-cc-pV5Z basis set using Gaussian 09 \cite{Gaussian}.
%References \cite{Ref5,Ref6,Ref7,Ref8,Ref9,Ref11,Ref12,Molecule1,Molecule2,Molecule3,Molecule4}
References \cite{Ref5,Ref6,Ref7,Ref8,Ref9,Ref11,Molecule1,Molecule2,Molecule3,Molecule4}
are all experimental data.
We find good agreements between the EQP energies of the forward ($\rightarrow$) electron absorption process 
and the backward ($\leftarrow$) electron release process.
Maximum deviation from the reference value is
1.0~eV in the case Be$^+$ $\leftarrow$ Be$^*$ for atoms, 
and 0.6~eV in the case Li$_2^+$ $\leftarrow$ Li$_2^*$ for molecules,
which are %the values calculated using
smaller than the deviations in the OQP theory 
%(values in the 
% in Table \ref{Table:EQP})
(values inside parentheses in Tables~\ref{Table:atoms} and \ref{Table:molecules});
3.2~eV in the case Al$^+*$ $\leftarrow$ Al. %the values calculated using
The average deviations are 0.4~eV and 0.3~eV, respectively, for atoms and molecules;
they are the same orders for both the OQP and EQP theories
%(values without parentheses in Talbe \ref{Table:EQP}).
(values outside parentheses in Tables~\ref{Table:atoms} and \ref{Table:molecules}).
%This means that
This means that there is no specific deviation in the EQP theory, %compared with the ordinary QP theory,
demonstrating the validity of the present theory. 
To obtain better agreement, it would be necessary to use the self-consistent GW($\Gamma$) methods \cite{Kuwahara,Kuwahara2}.
The resulting QP energy spectra of Li$^+*$ calculated by $G_0W_0$ are shown in Fig.~\ref{Fig1}.
It is seen from this figure that the agreement between the theory and experiment is very good already in the present approximation.

\section{Several Important Remarks}
\label{remarks}

Before ending this paper, we will give several important remarks.

(1) First of all, it is easy to see that the virial theorem holds for any excited eigenstate
of the $M$-electron Hamiltonian 
\begin{align}
H^M\{\br_i,R_k\} = -\frac{\hbar^2}{2m}\sum_i^M\nabla_i^2 + \frac{e^2}{4\pi\e_0}\left(
-\sum_i^M\sum_k\frac{Z_k}{|\br_i-\R_k|}
+\sum_{i>j}^M\frac{1}{|\br_i-\br_j|}
+\sum_{k>l}\frac{Z_kZ_l}{|\R_k-\R_l|} \right),
\label{Hamiltonian}
\end{align}
where $\{\br_i\}$ and $\{\R_k\}$ denote the positions of electrons and nuclei.
In fact, since the $M$-electron wave function $\Psi_{\g}^M\{\br_i\}$
of this excited eigenstate $\g$ is normalized as
%\begin{align}
$
\int\Psi^{M\ast}_{\g}\{\alpha\br_i\}\Psi_{\g}^M\{\alpha\br_i\}\prod_i^M d(\alpha\br_i)=1
$
%\end{align}
when all lengths are scaled by $\alpha$, the total energy of this excited eigenstate
at the equilibrium atomic positions $\{\R_k\}$,
\begin{align}
E^M_{\g}
=\int \Psi^\ast_{\g}\{\alpha\br_i\} H^M\{\alpha\br_i,\alpha\R_k\}
\Psi_{\g}\{\alpha\br_i\}\prod_i^M d(\alpha\br_i),
\end{align}
has the $\alpha$-dependence only in the form of
$\langle\hat{T}\rangle/\alpha^2+\langle\hat{V}_{\rm total}\rangle/\alpha$,
where $H^M\{\br_i,\R_k\}$ is given by (\ref{Hamiltonian}), $\langle\hat{T}\rangle$
and $\langle\hat{V}_{\rm total}\rangle$ denote
the kinetic and potential contributions to $E^M_{\g}$.
If the atomic configuration is fully relaxed at this excited state,
the total energy must be stationary with respect to the change in $\alpha$. Therefore,
differentiating $E^M_{\g}$ with respect to $\alpha$ and setting $\alpha$ at unity,
the virial theorem $2\langle\hat{T}\rangle+\langle\hat{V}_{\rm total}\rangle=0$
is readily derived.

\begin{figure}[hptb]
\begin{center}
\includegraphics[width=50mm]{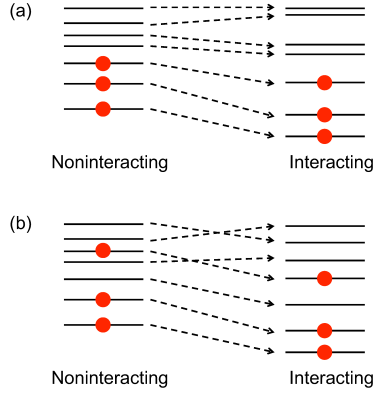}
\caption{(a) Adiabatic switch on process for the ground state:
Occupied QP levels must be below empty QP levels, {\it i.e.},
level crossing is not allowed between the occupied and empty QP levels.
(b) Gradual switch on process for an excited state:
QP levels can change the order,
{\it i.e.}, occupied and empty QP levels can intersect each other.}
\label{X}
\end{center}
\end{figure}

(2) Second, the adiabatic switch on process in the MBPT can be extended to excited eigenstates
as is written in Section 5-4 in Ref. \cite{Nozieres}.
In fact, Wick's theorem holds also for an excited eigenstate
if we transform creation and annihilation operators
from the coordinate representation to the one-particle state representation
of the corresponding noninteracting ($H^{(2)}=0$) system,
exchange the role of creation and annihilation operators
for occupied and empty states of the excited eigenstate,
and introduce the normal product as usual.
When the interaction $H^{(2)}$ is gradually switched on, the QP levels may apparently change the order.
However, this does not matter at all, because, in the case of an excited state, it is not necessary to require that
the noninteracting ground state moves onto the interacting ground state without level crossing.
Even if the order of occupied and empty levels changes during the switch on process, 
we can keep track of each QP state; see Fig. \ref{X}.
In the noninteracting system, the QP states refer to the one-particle states.
There is a one-to-one correspondence between the one-particle states of the noninteracting system
and the QP states of the interacting systems, although the orders of these states may be different.
There is no constraint on the orders of the one-particle states and the QP states,
and they can be different to each other.  
Therefore, we should not use the standard words ``the adiabatic switch on process''
for this process because the word ``adiabatic'' involves the meaning of no level crossing.
Instead, it would be better to use the words ``the gradual switch on process'' to avoid confusion.
The gradual switch on of the interaction from time $t=-\infty$ to $t=0$ transforms
the noninteracting excited state (at $t=-\infty$) into the interacting excited state (at $t=0$).

(3) Third, according to the Klein's derivation (see, for example, Section 5-6 in Ref. \cite{Nozieres}),
the total energy can be alternatively written as
\begin{align}
E^M_{\g}[G] - E^M_{\g}[G^{(0)}] = \Phi[G] + \Tr\left[\frac{G_s}{G^{(0)}_s} -1 - \ln \frac{G_s}{G^{(0)}_s}\right],
\label{E}
\end{align}
where
$G^{(0)}_s$ is the Green's function of the non-interacting system,
{\it i.e.}, the system described by the one-body part of the Hamiltonian $H^{(1)}$ only,
$\Tr=i\sum_s\int d\br\int_{-\infty}^{\infty}e^{i\eta\om}d\om/2\pi$
with $\eta=0^+$, and
$\Phi[G]$ is the
the Luttinger--Ward functional \cite{LW,Nozieres,Kuwahara}
\begin{align}
\Phi[G]=-\sum_{k=1}^{\infty}\frac{1}{2k}\Tr\int d\br'
\Sigma^{(k)}_s(\br,\br';\om)G_s(\br',\br;\om).
\end{align}
Here
$\Sigma^{(k)}_s(\br,\br';\om)$ represents
the $k$-th order skeleton diagram of the self-energy.
%In Eq. (\ref{E}), $\Tr [G_s/G^{(0)}_s - 1]$ is no other than $\Tr \Sigma_s G_s = -2E_{\rm int}$
In Eq. (\ref{E}), $\Tr [G_s/G^{(0)}_s - 1]$ is no other than $\Tr [\Sigma_s G_s] = -2E_{\rm int}$
(see Galitsukii--Migdal's formula (\ref{Galitskii}) and the Dyson equation (\ref{Dyson_w}) with
$G^{(0)\,-1}_s = \om - h^{(1)}_s$),
and $-\Tr\ln G_s/G^{(0)}_s$ can be evaluated as $\sum_{\mu}^{\rm occ}\e_{\mu} - E^M_{\g}[G^{(0)}]$
and $E^M_{\g}[G^{(0)}]=\sum_{\mu}^{\rm occ}\e^{(0)}_{\mu}$ with the eigenvalue of the non-interacting
system $\e^{(0)}_{\mu}$.
We have also the variational principles
$\delta\Phi[G]/\delta G_s %\propto
 = - \Sigma_s$
and
$\delta E^M_{\g}[G]/\delta G_s =0$.\cite{LW,Nozieres}

(4) The Ward--Takahashi identity \cite{Ward,Takahashi,WT},
\begin{align}
\delta(x-z)\,G^{-1}_s(z,y)
- G^{-1}_s(x,z)\,\delta(z-y)
=i\sum_{\alpha=0}^4\frac{\d}{\d z_{\alpha}}\Gamma^{(\alpha)}_s(x,y,z),
\label{WardTakahashi}
\end{align}
which is equivalent to the continuity equation for the electron density
(gauge invariance), holds also for the excited eigenstate.
Here $\Gamma^{(\alpha)}_s(x,y,z)$
are the scalar ($\alpha=0$) and vector ($\alpha=1,2,3$) vertex functions defined as
\begin{align}
\bra{\Phi_{\g}^M}T[j_{\alpha}(z)\psi_s(x)\psi^{\dagger}_s(y)]\ket{\Phi_{\g}^M}
=-\int G_s(x,x')\Gamma^{(\alpha)}_s(x',y',z)G_s(y',y)dx'dy',
\end{align}
where the four-current density $j_{\alpha}(x)$ is defined as
\begin{align}
j_0(x) & = \rho(x) = \sum_s\psi^\dagger_s(x)\psi_s(x),
\\
\j(x) & = - \rho(x)\A(x) % /c
 - \frac{i\hbar}{2m} \sum_s
\left\{\psi^{\dagger}_s(x)\nabla\psi_s(x)-[\nabla\psi^{\dagger}_s(x)]\psi_s(x)\right\}
\end{align}
with the vector potential $\A(x)$. % and the light velocity $c=137.036$ in the atomic unit.

(5) For the discussion of transport properties, it is important to guarantee
the macroscopic conservation laws of the total energy, the total number of electrons,
the total momentum, and so on.
These macroscopic conservation laws have been rigorously discussed
%for the nonequilibrium Green's functions
by Baym and Kadanoff \cite{BaymKadanoff,Baym} for the ground state.
Their argument may be generalized to the case of an arbitrary excited eigenstate,
and the macroscopic conservation laws hold for arbitrary excited states,
when the necessary symmetries suggested by %Baym and Kadanoff \cite{BaymKadanoff,Baym}
them
are satisfied for the excited-state Green's functions.
In general, however, we expect that the total number of electrons is conserved
in our excited-state theory because the local charge conservation is guaranteed
by the Ward--Takahashi identity (\ref{WardTakahashi}).
Moreover, we expect that the total energy and the total momentum of
the electronically excited many-atom systems should be conserved,
if the system is isolated.
% and $u(\br,t)$ is the Coulomb potential
% difference caused by the self-induced movements of nuclei.
This is simply due to the time-reversal symmetry and the translational symmetry
in space of the total Hamiltonian of the many-atom system.

(6) However, the fluctuation-discipation theorem is not guaranteed for arbitrary excited states
because it is based on the canonical ensemble, i.e. the thermal equilibrium,
at a given temperature.
Therefore, the linear response theory is not developed here and left for the future study.

(7) Last, we discuss how the DFT is modified for the excited state.
Similar to the ordinary DFT, we introduce a universal function
$F$ defined as
\begin{align}
F = \bra{\Phi_{\g}^M}H_{\rm eg}\ket{\Phi_{\g}^M},
\label{F}
\end{align}
where $H_{\rm eg}$ is the Hamiltonian of the interacting electron gas.
By definition, $F$ is a unique functional of the eigenstate $\ket{\Phi_{\g}^M}$
of the full Hamiltonian $H$.
This statement can be also straightforwardly deduced from the Luttinger--Ward theory,
because $F$ is expressed as a unique functional of the Green's function,
which is by definition (see Eq. (\ref{G1a})) a unique functional
of the eigenstate $\ket{\Phi_{\g}^M}$.
(Note that the rest contribution to the total energy is just the external potential $v(\br)$
coupled to the electron density $\rho(\br)=\sum_s\rho_s(\br)$.)
Below we show that this eigenstate $\ket{\Phi_{\g}^M}$
and therefore $F$ are unique functionals
of the electron density $\rho(\br)$ and the occupation numbers of all KS levels.
For this purpose, we use the {\sl reductio ad absurdum} along the same line as the Hohenberg--Kohn's theory \cite{HohenbergKohn}.
Let us assume that two different local potentials $v(\br)$ and $v'(\br)$ give the same electron density $\rho(\br)$,
and show that this assumption leads to a contradiction.
The Hamiltonians with the local potentials $v(\br)$ and $v'(\br)$ are written as $H$ and $H'$, respectively.
Their eigenvalues and eigenstates are written as $E^M_{\g}$, ${E'}^M_{\g}$, $\ket{\Psi_{\g}^M}$, and $\ket{{\Psi'}_{\g}^M}$,
which satisfy the eignvalue equations $H\ket{\Psi_{\g}^M}=E^M_{\g}\ket{\Psi_{\g}^M}$
and $H'\ket{{\Psi'}_{\g}^M}={E'}_{\g}^M\ket{{\Psi'}_{\g}^M}$.
The excited eigenstate is orthogonal
to all lower excited eigenstates as well as the ground state.
Therefore, we can still use the variational principle that $E^M_{\g}$ (or ${E'}_{\g}^M$) become minimum
when $\ket{\Psi_{\g}^M}$ (or $\ket{{\Psi'}_{\g}^M}$) is the true excited eigenstate
inside the Fock space orthogonal to all lower excited eigenstates and the ground state.
Then, inside this Fock space, we expect \cite{Ohno}
\begin{subequations}
\begin{align}
&{E'}^M_{\g} = \bra{{\Psi'}_{\g}^M}H'\ket{{\Psi'}_{\g}^M} < \bra{\Psi_{\g}^M}H'\ket{\Psi_{\g}^M}
\nonumber\\
& = \bra{\Psi_{\g}^M}H\ket{\Psi_{\g}^M} + \int [v'(\br)-v(\br)]\,\rho(\br) \, d\br
\nonumber\\
& = E^M_{\g} + \int [v'(\br)-v(\br)]\,\rho(\br) \, d\br
\end{align}
and
\begin{align}
&E^M_{\g} = \bra{\Psi_{\g}^M}H\ket{\Psi_{\g}^<} < \bra{{\Psi'}_{\g}^M}H\ket{{\Psi'}_{\g}^M}
\nonumber\\
& = \bra{{\Psi'}_{\g}^M}H'\ket{{\Psi'}_{\g}^M} + \int [v(\br)-v'(\br)]\,\rho(\br) \, d\br
\nonumber\\
& = {E'}^M_{\g} + \int [v(\br)-v'(\br)]\,\rho(\br) \, d\br.
\end{align}
\end{subequations}
Adding these two inequalities, we obtain
\begin{align}
E^M_{\g} + {E'}^M_{\g} < E^M_{\g} + {E'}^M_{\g}.
\end{align}
Thus we find that the initial assumption was wrong and that $v(\br)$ is a unique functional of $\rho(\br)$
when the corresponding state is constructed as the $\g$-th excited eignstate.
Since $v(\br)$ determines the form of the Hamiltonian $H$ uniquely,
its eigenstate $\ket{\Psi_{\g}^M}$ itself is a unique functional of the density $\rho(\br)$
and the information that this is the excited level $\g$.
Since the latter information is equivalent to the knowledge of the occupation numbers of all KS states,
we thus conclude that $\ket{\Psi_{\g}^M}$ and therefore $F$ are unique functionals of the electron density $\rho(\br)$
and the occupation numbers of all KS states only.
The variational principles still holds within the Fock space orthogonal to all lower excited states and the ground state,
which can be controlled by the choice of the occupation numbers of KS states.
Therefore only additional requirement in the extension of the ordinary DFT
to the electronic excited state is that we have to impose the constraint on the occupation numbers of all KS states.
In other words, the ordinary DFT still holds for the excited state
if we additionally give the information on the occupation numbers of all KS states.
This part only influences the exchange-correlation potential in the DFT.
We expect that there is a close similarity between the DFT for excited states and
the QP theory for initial excited eigenstates.
This argument encourages a direct use of the DFT to an electronic excited state such as
in \cite{MauriCar}, if a reliable exchange-correlation functional is used.

\section{Summary}
\label{summary}

In this paper, we have shown that the quasiparticle (QP) picture,
which correspond to the photoemission or inverse potoemission (PE/IPE) spectroscopy,
holds exactly for arbitrary excited eigenstate $\ket{\Psi_{\g}^M}$ of the system.
The QP energy spectra correspond to the PE/IPE energy spectra,
and the QP wave functions yields the electron density and the expectation values
of the kinetic and local potential energies.
They are the solution of the QP equations coupled with the equation for
the energy-dependent self-energy.
The interaction energy $E_{\rm int}$ of $\ket{\Psi_{\g}^M}$ eigenstate
is expressed by the energy-dependent self-energy and the QP wave functions
in Galitski--Migdal's form. 
All the formulation is given either with or without introducing the Green's function.
We have tried to drive the closed set of equations as simple as possible
to be readable for general readers who are not familiar to the field theoretical MBPT.
In this derivation, we showed that the energy-dependence of the self-energy is not an essential difficulty.
Some simple calculations of isolated atoms and molecules using the one-shot $GW$ approximation
demonstrate the validity of the present theory.
We also showed that the virial theorm, the Ward--Takahashi identity, and the conservation laws
hold and the Luttinger--Ward functional exists for arbitrary excited eigenstates.
Last, we gave a comment on the extension of density functional theory (DFT) to
arbitrary excited eigenstate.
We hope that the present theory makes a breakthrough toward the first-principles calculations
of arbitrary electronic excited states of materials.

\begin{acknowledgments}
We thank Steven G. Louie for helpful discussions.
This work has been supported by the grant-in-aid for Scientific Research
on Innovative Areas (Grant No.$\,$25104713) from MEXT and
the grant-in-aid for Scientific Research B (Grant No.$\,$25289218) from JSPS,
and also by the high performance computer infrastructure (HPCI)
(Project IDs.$\,$hp150231, hp150259, hp160234)
and the post "K" projects (priority issue 7 and germinating issue 1)
both promoted by MEXT.
\end{acknowledgments}

%\appendix

%\section{}
%\label{Appendix A}

%\section{}
%\label{Appendix B}

%\vspace*{5mm}
%\section{}
%\label{Appendix C}

%\newpage

%\bibliography{LGW}

%\begin{thebibliography}{}%

%\end{thebibliography}%

\end{document}